\documentclass[aps,pra,onecolumn,superscriptaddress,notitlepage,nofootinbib]{revtex4-1}
\usepackage{mathtools}
\usepackage{amsfonts}
\usepackage{latexsym}
\usepackage{relsize}
\usepackage{mathrsfs}

\DeclareMathAlphabet{\mathbbold}{U}{bbold}{m}{n}

\usepackage{euscript}%changes caligraphic font, see http://www.maths.usyd.edu.au/u/SMS/texdoc/euscript.pdf
\usepackage{amssymb}
\usepackage{graphicx}
\usepackage{amsmath}
\usepackage{amsbsy}
\usepackage[font=scriptsize]{caption}
\usepackage{subcaption}

\usepackage{amsthm}
\usepackage{bbm}
\usepackage{bm}
\usepackage{epsfig}
\usepackage{epstopdf}
\usepackage{dsfont}

\usepackage[colorlinks]{hyperref}
\usepackage[figure,table]{hypcap}
\usepackage{enumerate}
\usepackage{geometry}
\hypersetup{
	bookmarksnumbered,
	pdfstartview={FitH},
	citecolor={darkgreen},
	linkcolor={darkred},
       linktoc={page},
	urlcolor={darkblue},
	pdfpagemode={UseOutlines}}

\usepackage{color}

\definecolor{darkgreen}{RGB}{40,130,40}
\definecolor{darkblue}{RGB}{0,0,190}
\definecolor{darkred}{RGB}{238,0,0}
\usepackage{soul}
\usepackage{float}
\usepackage{algorithm}

\floatname{algorithm}{Protocol}

\def\EQ#1{\begin{eqnarray}#1\end{eqnarray}}
\newcommand{\djj}{d\kern-0.4em\char"16\kern-0.1em}

\newcounter{lem}

\newtheorem{prop}{Proposition}\def\PRO{\begin{prop}}\def\ORP{\end{prop}}
\newtheorem{coro}{Corollary}\def\COR{\begin{coro}}\def\ROC{\end{coro}}
\newtheorem{theo}{Theorem}\def\TH{\begin{theo}}\def\HT{\end{theo}}
\def\TH{\begin{theo}}\def\HT{\end{theo}}
\newtheorem{defi}[prop]{Definition}\def\DE{\begin{defi}}\def\ED{\end{defi}}
\newtheorem{lemme}[lem]{Lemma}\def\LE{\begin{lemme}}\def\EL{\end{lemme}}

\def\dm#1{\left|#1 \right\rangle \left\langle #1 \right|}

\usepackage{amsmath,amsfonts,amssymb}
\usepackage{wrapfig}
\usepackage{graphicx}
\usepackage{bbm}
\usepackage{graphics}

\def \beq {\begin{equation}}
\def \eeq {\end{equation}}
\def \ba {\begin{eqnarray}}
\def \ea {\end{eqnarray}}

\input{Qcircuit}

\begin{document}

\title{Framework for learning agents in quantum environments}
\date{\today}
\author{Vedran Dunjko}
\email{vedran.dunjko@uibk.ac.at}
\affiliation{Institut f\"{u}r Theoretische Physik, Universit{\"{a}}t Innsbruck, Technikerstra{\ss}e 25, A-6020 Innsbruck, Austria}
\affiliation{Division of Molecular Biology, Ru\djj er Bo\v{s}kovi\'{c} Institute, Bijeni\v{c}ka cesta 54, 10002 Zagreb, Croatia.}
\author{Jacob M. Taylor}
\email{jmtaylor@umd.edu}
\affiliation{Joint Quantum Institute, National Institute of Standards and Technology, Gaithersburg, MD 20899 USA} 
\affiliation{Joint Center for Quantum Information and Computer Science, University of Maryland, College Park, MD 20742 USA}
\author{Hans J. Briegel}
\email{hans.briegel@uibk.ac.at}
\affiliation{Institut f\"{u}r Theoretische Physik, Universit{\"{a}}t Innsbruck, Technikerstra{\ss}e 25, A-6020 Innsbruck, Austria}

\begin{abstract}
In this paper we provide a broad framework for describing learning agents in general quantum environments. We analyze the types of classically specified environments which allow for quantum enhancements in learning, by contrasting environments to quantum oracles.
We show that whether or not quantum improvements are at all possible depends on the internal structure of the quantum environment.
% We show that, in general, quantum improvements are not possible if classical environments are straightforwardly generalized to full quantum environments. 
If the environments are constructed and the internal structure is appropriately chosen, or if the agent has limited capacities to influence the internal states of the environment, we show that improvements in learning times are possible in a broad range of scenarios. Such scenarios we call luck-favoring settings. 
The case of constructed environments is particularly relevant for the class of model-based learning agents, where our results imply a near-generic improvement.
\end{abstract}
\vspace{9cm}

\maketitle
%\tableofcontents

\section{Introduction}
\label{intro}
The capacity to learn, and to adapt to unknown environments is one of the quintessential characteristics of intelligent agents.
The development of learning models has, consequently, been one of the central areas of research in the broad field of artificial intelligence (AI) from its very beginnings.
Nonetheless, the central goal of AI, the design of autonomous, intelligent learning agents is still far from realized. In contrast, more computation-orientated aspects of so-called applied AI, for instance machine learning, classification, pattern recognition, even game playing (e.g. computer chess), have continuously progressed over the years, despite occasional set-backs.
 
The field of quantum information processing (QIP), which has had a profound impact on computer and communication sciences, has also recently started directly influencing such applied aspects of AI. For example, the theory and algorithms for classification and clustering, in both supervised and unsupervised settings, and which consider and exploit quantum mechanics, have been provided \cite{2013_Lloyd, 2013_Aimeur,2002_Sasaki,2004_Servedio, 2008_Neven, 2012_Neven}. 

Other works have exploited quantum mechanics to reduce space or time complexity of particular reinforcement learning approaches \cite{2005_Dong, 2014_Paparo}, when the environment is classical. 

However, no general framework for learning from experience (often formalized as reinforcement learning \cite{SuttonBarto98}), where an agent and an environment \emph{interact quantum-mechanically}, has been presented so far. Moreover, the potential of utilizing quantum mechanics in reinforcement learning scenarios with a genuine quantum access to environments has been thus far mostly unexplored.

Here, we provide the first steps in this direction. We begin by specifying a framework for classical learning from experience which subsumes the standard setting of computational reinforcement learning.
We then consider the consequences when the agent, environment and the interaction become fully quantum mechanical. 

In particular, we explore what it even means to learn in a setting where the agent, and the environment become, for instance, entangled.
As the central concept, from which learning figures of merit are derived, we posit the \emph{history of the interaction} between an agent and the environment.
In the classical setting, the history is unambiguously defined, however, in the quantum case, due to the superposition principle and entanglement, the history may be completely undetermined\footnote{
Some of the aspects of the notion of a history  (or rather the problems with the notion of history in a full quantum setting) we introduce here may be reminiscent to ideas in certain formulations of quantum mechanics, e.g. the consistent histories approach \cite{1984_Griffiths}, or the related decoherent histories approach \cite{1990_Gell-Mann}. Our notion of history, however, is not intended to address any foundational questions regarding quantum mechanics.}. To resolve this issue, we introduce the notion of \emph{a tester} - a third entity, which can monitor the interaction, and which, ultimately, decides whether the agent was successful in learning. {We consider a broad class of testers and, as a first result, show that \emph{classical testers,} that is testers which record the entire history of interaction verbatim, prohibit any quantum speed-up. Nonetheless, other types of testers, specifically \emph{sporadic testers}, which allow periods of interaction to proceed unmonitored, may allow improvements.}

The basic question we then ask is, can we, given a classical agent interacting with a classical task environment, construct a quantum agent, which utilizes quantum interaction with the \emph{same} (in a sense which we specify later) environment, and which performs better, {relative to a meaningful tester.}

In our analysis, we draw parallels between oracular models of computation, and agent-environment interaction settings, and show that only very simple environments correspond to standard oracles. 
The most interesting and challenging environments, from the perspective of learning, are those which have memory themselves, and in which the state of the environment depends on the history of the interaction. This memory is typically unaccessible to the agent - the outputs of the agent are typically not returned, but lost to the environment. This prohibits standard quantum information techniques to be of direct use: superposition inputs are lost, or de-phased, and similar holds for entangled inputs.
We present two methods of circumventing these problems.

The first method is relevant for computational settings (and also in model-based approaches to learning
%\footnote{In model-based approaches to learning, the agent, internally, constructs a map - a representation of the environment, which is updated through the agents experience, and which is used as a tool in planning, and decision making \cite{2003_Russel}.}
 which we address later), where the task environments (or the models of the environment) are often constructed. In such a setting, if the interaction between the agent and environment is appropriately tweaked, improvements in learning times become possible. 
{More specifically, for a given classical specification of a task environment $E$, it is possible to construct an `oracularized' variant of the environment $E_{oracle},$ which, in particular, returns all the relevant registers to the agent. Such environments are a generalization of standard oracles utilized in QIP. Oracularized environments, however, do not recover the same behavior as the original environment $E,$ and would be useless for a classical agent. To enable a fair comparison between classical and quantum agents, we then define controllable environments $E_{control}$ which the agent can access in their original ($E$) or oracularized ($E_{oracle}$) instantiation on demand.}
 
 As our main example we then show that, {in the case the agent has access to such a controllable environment,}  the \emph{exploration phase} of learning, which precedes the \emph{exploitation phase}, can be generically quadratically sped-up by using a quantum interaction (effectively by exploiting Grover's search), {under the presence of a sporadic tester. }
 This, in turn, allows us to further show that the actual overall learning performance of an agent, in many cases, is also significantly improved.

A second method for resolving the problems occurring in quantum environments with memory, which does not resort to explicitly constructing the environment, resorts to moderate manipulation of the memory of the environment itself.
To facilitate this, we introduce additional powers to the agent, \emph{register hijacking} and \emph{register scavenging}, which in essence allow the agent to modify certain `memory components' of the environment.
{More specifically, register scavenging implies that any register which is `traced out' by the environment (that is, a register the environment no longer needs to control) is assumed to then be fully accessible for the agent. Register hijacking is more intrusive, and assumes that the agent can alter parts of the active registers of the environment. Nonetheless, neither hijacking nor scavenging change the specification of the environment.}
 Using these additional options for the agent, we achieve the same results as in the constructed environment setting. 

{Neither hijacking nor scavenging are, in general, realistic options for the agent. 
Nonetheless, there are a few settings where they could be realized. For instance, if the `quantum environment' is realized by a controlled part of a quantum experiment, then it is conceivable that a well-designed agent may have access to the relevant registers\footnote{This is possible even when some of the details of the actual dynamics driving the evolution of those subsystems may be unknown -- these are the cases when there is still something left to be learned. For instance, in metrological settings, one often assumes a general form of an unknown dynamics, but the specific parameters are unknown.}. }
 {On the other hand, in the context of model-based approaches to learning we can safely assume full control of the constructed (internal) environment.}
{Thus, the results obtained through scavenging and hijacking can also be applied to all model-based approaches to learning, where the representation of the environment is explicit. A possible advantage of the hijacking/scavenging approach, over an explicit construction of oracularized instantiations of the environment, is that the former can be applied generically, that is, without having to consider the details of a given model-based learning model.}

The outline of the paper is as follows. 
We begin in Section \ref{interaction} by introducing the classical and quantum notions of agent-environment interaction and learning, and establish the basic results pertaining to the quantum case.
In Section \ref{QAEI} we contrast environments to oracles, and consider types of environments where  generic improvements in learning performance may be possible, by utilizing full quantum interaction.
%investigates the quantum analogs of the classical agent-environment interaction setting. Here, we establish the basic results which arise when one allows quantum access to otherwise classically specified environments. 
Sections \ref{main-section} and \ref{lfa} contain our main results. There we characterize the types of constructed environments where we can provide generic improvements in learning performance, by utilizing full quantum interaction.  
In Section \ref{extensions} we further generalize the types of constructed task environments in which learning speed-ups are possible, and in Section \ref{indirect} we investigate the settings where unconstructed task environments allow for the same type of performance as constructed environments.
Section \ref{mod-based} provides an immediate application of our results in the context of so-called model-based learning agents.
We summarize our results in Section \ref{discussion}  and consider further lines of investigation.
In the Appendix (Section \ref{appendix}) we provide the remaining proofs of lemmas from the main text.

\section{Agent-environment interaction}
\label{interaction}
The basic concepts pertaining to classical agent-environment interaction are comparatively straightforward, and we present them first. In the process we also establish the notation for this paper. Following this, we briefly reflect on the minimal conditions for an agent to be considered a learning agent, and then we focus on the main topic of this section - a coherent definition of quantum agent-environment interaction in which notions of learning can be defined.

\subsection{Classical agent-environment interaction}
\label{class-inter}
Here we define what constitutes, for the purposes of this work, a classical agent-environment interaction. 
The basic components are the percept $(\mathcal{S})$ and action  $(\mathcal{A})$ sets which specify the possible outputs of the environment, and the agent, respectively.
We restrict our setting to synchronous, finite setting, meaning that the action and percept spaces are finite, and the interaction between the agent and the environment is turn-based\footnote{Turn-based implies that the agent and environment interact in discrete time steps, alternating between the agent and the environment.}.
These assumptions are common in standard AI literature, and are important for the quantum setting we present next.  
A realized interaction up to time step $t$, between the agent and the environment, that is a sequence $h_{t} = (s_1, a_2, s_3,s_4, \ldots, s_{t-1},a_{t}), s_i\in \mathcal{S}, a_j \in \mathcal{A}$ of alternating percepts and actions is called \emph{the $t-$step history} of interaction. With $\mathbf{H} = \bigcup_{t\geq 0} \mathbf{H}_t$ we will denote the set of all histories where  $\mathbf{H}_t$ denotes the set of all histories of length $t$.

The agent and environment are formalized as stochastic maps with memory, sometimes referred to as \emph{random systems} \cite{2002_Maurer}. 

At the $t^{th}$ time-step, and given the elapsed history $h_{t-1}$ the behavior of the agent (the environment) is characterized with the corresponding instantaneous maps, respectively:
\EQ{
M_{A}^{h_{t-1}}(s\in\mathcal{S} ) \in distr(\mathcal{A});\\
M_{E}^{h_{t-1}}(a\in\mathcal{A} ) \in distr(\mathcal{S}),
}
where $distr(\mathcal{X})$ denotes the set of probability distributions over the set $\mathcal{X}$. The superscripts denote the realized history up to time step $t-1$. The argument of the maps specifies the current percept (action) which the agent (environment) perceives. The output of the agent (environment) is an action (percept) sampled from a distribution over the action (percept) space. To exemplify, 
the agent outputs, at time step $t$, given percept $s$ and history $h_{t-1}$, the action $a$ which is sampled from the distribution $M_{A}^{h_{t-1}}(s)$.
The agent, and the environment, are then specified by the sequences of these characteristic maps $\{M_{A}^{h} \}_{h}$, $\{M_{E}^{h}\}_{h},$ indexed over the set of histories.

Given an agent, and an environment, due to the, in general stochastic, nature of the characteristic maps, particular histories can occur with varying probabilities. We will with $A\leftrightarrow_t E$ denote the probability distribution over $t-$step histories, and with $A\leftrightarrow E$ the distribution over all histories $\mathbf{H}$, realized by the agent $A$ and environment $E$. We will also refer to $A\leftrightarrow_t E$ as \emph{the interaction} between the agent $A$ and environment $E$. 
As a technicality, we need to standardize the initial conditions of an interaction. We will, as a convention, assume that the interaction begins with the environment outputting the first percept.

To formalize this, we will assume that the action and percept spaces also contain the empty percept/action element $\epsilon$, which is also the first element of any history, and the first percept of any interaction. Then, given an agent and and environment, the first action output is given with $a \sim \mathcal{M}^{\epsilon}_A(\epsilon)$, followed by the environment's step  $s\sim \mathcal{M}^{\epsilon}_E(a)$. With $x \sim d\in distr(\mathcal{X})$ we have denoted that the element $x \in \mathcal{X}$ is distributed according to the distribution $d$ over the state space $\mathcal{X}$.  If we require the agent to be on the move first, we will simply assume that the first percept is the empty percept $\epsilon$. 

The formal definition of the interaction is best given recursively. The distribution $A \leftrightarrow_2E$ is specified with $P(A \leftrightarrow_2 E = a) = P(\mathcal{M}^{\epsilon}_A(\epsilon) = a),$ where we understand the characteristic maps as random variables. The indexing of the interaction starts with 2, as the first move of the environment is defined to be the trivial percept $\epsilon$.

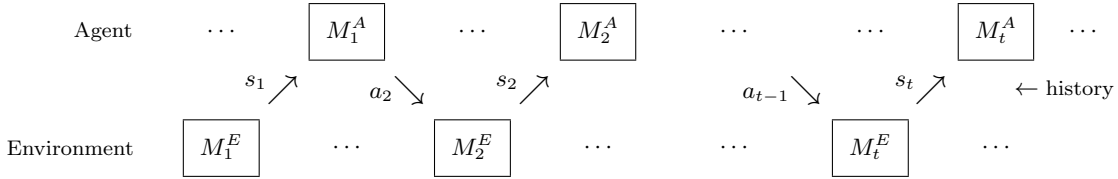
\begin{figure}[!h]
\begin{equation}
\hspace{-15pt} \Qcircuit @C=2em @R=1.2em {
\lstick{\hspace{0.7cm} {{}}}    & \push{ {\footnotesize \textup{\hspace{22pt} Agent} } } & \cdots        &\push{ \fbox{\raisebox{0pt}[10pt][5pt]{} ${M}^A_1$  }         } & \cdots                           &\push{ \fbox{\raisebox{0pt}[10pt][5pt]{} ${M}^A_2$  }         }  & &\cdots   &  & \cdots &\push{ \fbox{\raisebox{0pt}[10pt][5pt]{} ${M}^A_t$  }         }& \cdots &    \\
\lstick{}                    &   &       &     \hspace{-60pt}    \raisebox{0.5em}{ $s_1$}   \mathlarger{\mathlarger{\mathlarger{\nearrow} }}         &     \hspace{-60pt}    \raisebox{0em}{ $a_2$}   \mathlarger{\mathlarger{\mathlarger{\searrow} }}   &   \hspace{-60pt}    \raisebox{0.5em}{ $s_2$}   \mathlarger{\mathlarger{\mathlarger{\nearrow} }}        &  &  &&         \hspace{-70pt}    \raisebox{0em}{ $a_{t-1}$}   \mathlarger{\mathlarger{\mathlarger{\searrow} }}  &   \hspace{-60pt}    \raisebox{0.5em}{ $s_t$}   \mathlarger{\mathlarger{\mathlarger{\nearrow} }}     &  \hspace{-0.5cm}\leftarrow \textup{\footnotesize{history}} &&\\
\lstick{}            &   \push{{\footnotesize \textup{Environment}}}         &       \push{ \fbox{\raisebox{0pt}[10pt][5pt]{} ${M}^E_1$  }         }       & \cdots                     & \push{ \fbox{\raisebox{0pt}[10pt][5pt]{} ${M}^E_2$  }         }   &  \cdots         &                                                           &\cdots   & & \push{ \fbox{\raisebox{0pt}[10pt][5pt]{} ${M}^E_t$  }         }  &            \cdots                &       & &
}\nonumber
\end{equation}
\caption{\label{class-fig} Classical agent-environment interaction. On an abstract level, any turn-based agent-environment interaction can be visualized as an exchange of messages between the agent and the environment. The maps of the agent, and environment, at each time-step may depend on the entire histories, here for simplicity denoted just by the time-step designation in the subscript. }
\end{figure}

The interactions of even length (i.e. ending with the agent's move) are specified with
\EQ{P(A \leftrightarrow_{2t} E = h_{2t}) =  P\left(A \leftrightarrow_{2t-1} E = {\left[h_{2t}\right]}_{-1}\right)P\left(\mathcal{M}^{ {\left[h_{2t}\right]}_{-2} }_A\left({\left[h_{2t}\right]}_{2t-1}\right) =  {\left[h_{2t}\right]}_{2t}\right),
}
where ${\left[h_{2t}\right]}_{-2}$ and ${\left[h_{2t}\right]}_{-1}$ denote the history of length $2t-2$ and and history of length $2t-1$, obtained by dropping the last two elements, and the last element from $h_{2t}$, respectively.  ${\left[h_{2t}\right]}_{2t}$ denotes the last element of the same history. The interactions of odd lengths are defined analogously. An illustration of a classical turn-based agent-environment interaction is given in Fig. \ref{class-fig}.

In reinforcement learning, one standardly introduces a notion of a \emph{reward} $\lambda$, from a reward set $\Lambda$, which is issued by the environment to the agent after each action. The reward, intuitively, specifies whether a performed action, in the given context, was correct\footnote{In a broader framework, the origin of the rewards may not be explicitly the environment. For instance, whether or not some sequence of percepts was `good' for the agent may be generated internally within the agent - think sensors on a robot which detect that the position of a robotic arm is becoming harmful for the arm.  This `pain signal' then constitutes a negative reward. Sometimes, this is modeled by the introduction of a third entity, often called \emph{the critic}, which is situated within the agent, and produces the rewards from the history. Since the issuing, or non-issuing of the reward is always dependent on the responses of the environment (otherwise, the agent is `hallucinating') the true origin of the reward is not important for our considerations. }.
In our setting, we will not explicitly model the rewarding step, and will assume that the percept space contains a reward-specifying component, so $\mathcal{S} =\mathcal{S}' \times \Lambda,$ and where $\mathcal{S}' $ is the `raw' percept space.

In the context of learning, one typically assesses the performance of a learning model (agent), relative to some task environment, by tracking the interaction between the agent and the environment. As an example, the figure of merit of interest  is often the learning efficiency, that is the probability of the agent receiving a reward at time-step $t$, as a function of $t$.
In our model, a figure of merit $f$ (also called \emph{efficiency measure}) is, essentially, any function defined on the set of histories $\mathbf{H}$. 
In this paper, we will assume that the codomain of this function is the set of real numbers (any totally ordered set would do, as our goal is to compare the performances of environments).
Most of the time, one is interested in the average behavior of the agent, for which case we will assume the function $f$ is extended by convex-linearity, and defined on the space of interactions $A\leftrightarrow E$.\\

%
%We define the \emph{characteristic map of the agent (environment)} with:
%\EQ{
%M_{A} : S \cup\bigcup_{i=0}^{\infty}(\mathcal{S} \times \mathcal{A})^{\times i} \rightarrow \textup{distr}(\mathcal{A}),
%}
%\EQ{
%M_E : \{\epsilon \} \cup  \bigcup_{i=0}^{\infty}(\mathcal{S} \times \mathcal{A})^{\times i} \rightarrow \textup{distr}(\mathcal{S}),
%}
%We introduce the notion of history $h_n$ which is a particular sequence of (percept,action) pairs, of length $n$, and the distribution over histories $A \leftrightarrow E$ which is realized by an agent-environment interaction. 
%Given the notion of an $n-$step history, we can recast the definition of the agent (and, analogously for the environment) which emphasize the notion of memory, and for which we will give the quantum analogs. The $n^{th}$ step instantaneous map of the agent ${M_{A}}^{h_{n-1}}$, given history $h_{n-1}$ is defined as:
%\EQ{
%{M_{A}}^{h_{n-1}}(s\in\mathcal{S} ) = M_{A}(h_{n-1} \| s),
%}
%where $h_{n-1} \| s$ denotes the concatenation of the history with the percept $s$. In the representation above $h_{n-1}$ can be viewed as the contents of the memory of the agent.
%
%
%\red{Introduce the notion of memory, but only as an implementation-level statement, clarify it will be important in the quantum framework.}

\paragraph*{What it means to be a learning agent\\ }
%\vdc{This section is still a bit problematic, unclear what I want to say here... I start with classical notions, then go into quantum improvements...also it is too long, and rerereiterates what we do.}
\label{learningagent}
For the purposes of this work, we will adhere to the bare minimum of what it may mean to be a learning agent. In particular, we will just demand that the behavior (instantaneous maps) of the agent depends on the history of the agent. Concisely, an entity is a learning agent if its behavior can change under environmental stimulus.
While the capacity to change under stimulus may be necessary for learning, it certainly does not suffice for any notion of a \emph{good} learning agent. Indeed, various results in optimization and machine learning \cite{1996_Wolpert,1997_Droste}  cumulatively dubbed ``no free lunch theorems'' suggest that a robust definition, or an order on the set of learning agents which classifies how well they can learn, independent from a task environment, may be too much to ask for\footnote{We point out that the results we cite, and their interpretation, are not without controversy, see \cite{NFLorg}  for more information. However, at the time of writing of this work, to our knowledge, there is certainly no consensus on such an environment-independent classification of learning agents, and we do not presume to clarify this issue here.}.
Nonetheless, at least relative to fixed families of task environments, learning agents can be compared. In this sense, the approach we will use in this paper will ask whether a given classical agent can be `upgraded' to a quantum agent, and yield a relative improvement. We will show that, under certain assumptions, this can be done for a broad class of typical environments. We further discuss what learning problems, and what learning may be in Section \ref{main-section}.

\subsection{Quantum agent-environment interaction}
\label{q-Inter}
As the first formal step, percepts are represented as orthogonal basis states of the percept Hilbert space  $\mathcal{H}_\mathcal{S}~=~\textup{span} \{ \ket{s} \vert s \in \mathcal{S}  \}$. Analogously, for the action space we have $\mathcal{H}_\mathcal{A} = \textup{span} \{ \ket{a} \vert a \in \mathcal{A}   \}$, also satifying  $\bra{a}{a'}\rangle = \delta_{a,a'}$, where $\delta$ is the Kronecker-delta function.
To both the agent, and the environment, we assign internal memory:  finite (but arbitrarily sized) registers $R_A$ and $R_E$
capable of storing histories, so with Hilbert spaces of the form $\mathcal{H}_\mathcal{A} \otimes\mathcal{H}_\mathcal{S} \otimes \mathcal{H}_\mathcal{A}  \cdots$.

Next, we specify the interface of the agent and the environment - that is parts of the system of the agent (environment) to which they both have access, in contrast to $R_A$ ($R_E$) which are reserved for the agent  (environment) exclusively.
There are two natural ways of defining the interface. In the spirit of quantum communication theory, we can define the unique common communication register $R_C$, with associated Hilbert space $\mathcal{H}_C$ sufficient to represent both actions and percepts, thus $\mathcal{H}_C~=~\{ \ket{x} | x\in \mathcal{S} \cup \mathcal{A} \}$. We assume actions and percepts are mutually orthogonal, so $\mathcal{H}_C$ is isomorphic to $\mathcal{H}_\mathcal{S} \oplus \mathcal{H}_\mathcal{A} $.

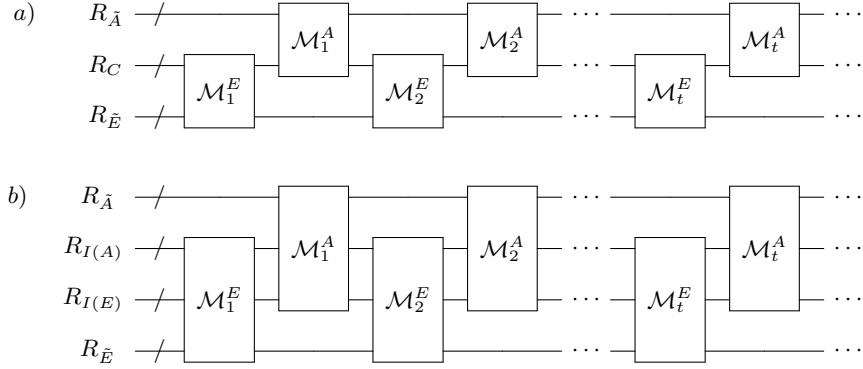
\begin{figure}
\begin{equation}
 \Qcircuit @C=1em @R=1.2em {
\lstick{a)\hspace{0.7cm}  R_{\tilde{A}}}  &{/} \qw&\qw                                 & \multigate{1}{\mathcal{M}^A_1}            & \qw                            & \multigate{1}{\mathcal{M}^A_2}                    & \qw &\cdots   &  & \qw & \multigate{1}{\mathcal{M}^A_t} &\qw &\cdots    \\
\lstick{R_C}                                  &{/} \qw & \multigate{1}{\mathcal{M}^E_1}  &\ghost{\mathcal{M}^A_1  }                     & \multigate{1}{\mathcal{M}^E_2}&\ghost{\mathcal{M}^A_2  }   &  \qw                          &\cdots  &   & \multigate{1}{\mathcal{M}^E_t}&\ghost{\mathcal{M}^A_t  }          &\qw &\cdots   &  \\
\lstick{R_{\tilde{E}}}                                  &{/} \qw & \ghost{\mathcal{M}^E_1  }           & \qw                          & \ghost{\mathcal{M}^E_2 }    &  \qw         &\qw                                                               &\cdots   & & \ghost{\mathcal{M}^E_t  } &  \qw                           & \qw              & \cdots &
}\nonumber
\end{equation}

\begin{equation}
 \Qcircuit @C=1em @R=1.2em {
\lstick{b)\hspace{0.7cm}  R_{{\tilde{A}}}\ }    &{/} \qw& \qw                                  & \multigate{2}{\mathcal{M}^A_1}        & \qw   & \multigate{2}{\mathcal{M}^A_2}                                  & \qw &\cdots   &  & \qw  & \multigate{2}{\mathcal{M}^A_t}                              &\qw  &\cdots    \\
\lstick{R_{I(A)}}                                   &{/} \qw  & \multigate{2}{\mathcal{M}^E_1} &\ghost{\mathcal{M}^A_1  }                     & \multigate{2}{\mathcal{M}^E_2}   &\ghost{\mathcal{M}^A_2  }            &  \qw &\cdots  &   & \multigate{2}{\mathcal{M}^E_t} &\ghost{\mathcal{M}^A_t  }           &\qw &\cdots   &  \\
\lstick{R_{I(E)}}                                   &{/} \qw  & \ghost{\mathcal{M}^E_1}            &\ghost{\mathcal{M}^A_1  }                     & \ghost{\mathcal{M}^E_2}   &\ghost{\mathcal{M}^A_2  }                   &  \qw &\cdots  &  & \ghost{\mathcal{M}^E_t}  &\ghost{\mathcal{M}^A_t  }                                &\qw  &\cdots     &  \\
\lstick{R_{\tilde{E}}\ }                                       &{/} \qw   & \ghost{\mathcal{M}^E_1}           & \qw                                         & \ghost{\mathcal{M}^E_2 }    &  \qw                                       & \qw & \cdots      && \ghost{\mathcal{M}^E_t  } & \qw                                    & \qw & \cdots &
}\nonumber
\end{equation}
\caption{\label{fig1}
Interaction between an agent and an environment. In case a) there is a unique communication register $R_C$, visible to both the agent and the environment, and this representation is easily comparable to Fig. \ref{class-fig}. In case b) the agent and the environment each have an interface register, visible to the other party. 
In the representation above, since we are using the dilated form of the maps, the registers $R_{\tilde{A}}$ and $R_{\tilde{E}}$ are assumed to contain the internal registers $R_{A},$ $R_{E}$, respectively, but also the purifying registers, if the characteristic maps are considered in a dilated form.
}
\end{figure}

A slightly more general definition, in the spirit of embodied cognitive sciences and robotics, defines the agent's and the environment's interfaces separately\footnote{In embodied cognitive sciences and robotics, the agent is an embodied entity, equipped with \emph{sensors} and \emph{actuators}, using which the agent perceives, and acts on its environment, respectively. Form this perspective, the physical degrees of freedom of the sensors (actuators) correspond to the register $R_{I(E)}$ ($R_{I(A)}$) . }. The agent  then comprises the internal register $R_A,$ and the interface register $R_{I(A)},$with Hilbert space $\mathcal{H}_\mathcal{A}.$
 The environment has, along with the internal register $R_E$ the interface register $R_{I(E)},$ with Hilbert space $\mathcal{H}_\mathcal{S}.$
The agent (environment) is then specified by sequences of completely positive trace preserving (CPTP) maps  $\{\mathcal{M}^{A}_i \}_i$ ($\{\mathcal{M}^{E}_i \}_i$) acting on the concatenated registers $R_AR_{I(A)}R_{I(E)}$ ($R_{I(A)}R_{I(E)}R_{E}$).
It will sometimes be useful to dilate the maps above to unitary maps, by using appended registers added to $R_A$ and $R_E$ if needed.
An agent-environment interaction is then specified by the sequential application of the maps, illustrated in Fig. \ref{fig1} for both specifications of the interface.

Unless otherwise specified, we will always assume that the initial state of the registers $R_A R_C R_E$ is a predefined fixed pure state, which is separable with respect to the three registers, and an analogous assumption is made in the embodied picture\footnote{Assuming a classically correlated state can be used to model prior knowledge the agent may have about the environment, and entangled initial states may give even more options for the agent and the environment. However, we will not be considering such cases in this work explicitly.}. Since the initial state is pure and separable, the state is also in product form with respect to the three partitions. The specific choice of the initial product state is irrelevant as the preparation of any particular choice may be subsumed into the initial maps of the agent and environment.

\subsubsection{Classical limit of the quantum setting}
\label{Class-limit}
We now discuss how the  quantum framework fits in the more usual classical framework of agents~\cite{2003_Russel}, specified in Section \ref{class-inter}.
For this we first define the notion of a classical agent (and analogously, a classical environment).
In the following, we will call any state, which is a tensor product of percept/action basis states, \emph{a classical state}. Probabilistic mixtures of such states (that is, states whose density operators are convex combinations of the corresponding projectors) are also classical states, and no other states are classical\footnote{{For completeness we note that classical mixed states are defined relative to a register/system under consideration: for instance, a Bell-pair state of two qubits is not classical, whereas the reduced states of both individual qubit are, as they are equiprobable mixtures of any two orthogonal states.}}.
The definition of classical states is analogous to the standard concept of {computational basis states} in quantum computing. The particular choice of such a basis will, in practice, depend on the particular systems forming the agent and the environment.

We give the following definitions for the communication-oriented model, where there is one interface register $R_C$.
Analogous statements are obtained for the embodied model, by substituting the register $R_C$ with the pair of registers $R_{I(A)}R_{I(E)}.$

\DE
The agent $A$ is \emph{classical}, if for every map $M \in \{ \mathcal{M}^{A}_k \}_k$ acting on $R_AR_C$ the following holds:  if the state of the register $\rho_{R_AR_C}$ is of the product form $\ket{\Psi} = \ket{\psi}_{R_{A}} \ket{s}_{R_{C}},$ where $\ket{\psi}_{R_{A}}$ is a classical state, and $\ket{s}_{R_{C}}$ is a classical percept state, then 
\EQ{
M(\dm{\Psi}) = \sum_{i,j} p_{i,j} \dm{\psi^i}_{R_A} \otimes \dm{a_j}_{R_C},
}
where $\ket{\psi^i}$ are classical states, $\ket{a_i}$ are classical action states and all $p_{i,j}$ are real.
\ED
In other words, the agent $A$ is classical if its maps neither generate entanglement, nor coherent superpositions of classical states, when acting on classical states\footnote{Note the in the definition above we refer to the defining maps of the agent, not their dilations - the purifying systems (which are not parts of the agents or environments memory) could, naturally, be entangled to both the registers $R_A$ and $R_C$.}.
A classical environment is defined analogously.

The setting of a classical agent which interacts with a classical environment constitutes a natural starting point for a sensible definition of a classical interaction. However, for the interaction itself, the internal states of the agent/environment should not matter. Thus we give the following, more general definition of classical interaction and we formally justify this choice later in this section.

\DE
\label{def-class-int}
The interaction between the agent $A$ and the environment $E$ is called \emph{a classical interaction} if at every stage of the interaction, the state of the combined registers can be represented in the form
\EQ{
\rho_{R_A R_C R_E} = \sum_{i} p_i\, \eta^i_{R_A} \otimes \rho^i_{R_C} \otimes \sigma^i_{R_E}
}
where each $\rho^i$ is a convex mixture of classical states, all $\eta^i$ and $\sigma^i$ are unit trace density matrices, and $p_i$ are (real) probabilities.
\ED
Given definition \ref{def-class-int}, a classical interaction between an agent and the environment does not entail that the agent and the environment are internally in classical states. However, we do prohibit entanglement between the the registers $R_A, R_C$ and $R_E$, and also coherent superpositions in the interface registers.

Next, we consider what, in the context of quantum agent-environment interactions, a proper analog of a history should be.
In the cases where, for instance, the states of the agent, environment and the interface are entangled, there is no straightforward analog of the classical history. Intuitively, since we are dealing with quantum systems, the history should be an observable of the systems.
 More precisely, it should surmount to a sequence of observables, defined for all the time steps of the interaction.
We formalize and and characterize a quantum history more broadly by introducing a third entity - a \emph{tester}. 

The tester is a sequence of CPTP maps $\{ \mathcal{M}^T_k \}$ which act on an external register $R_T$ and the communication register $R_C$ (analogously $R_{I(A)}R_{I(E)}$). Often we may assume that the maps are unitary (by dilating the maps, if needed).
In the \emph{tested} interaction between an agent and the environment, the map of the tester is applied after each map of both the agent and the environment.
The tester is not meant to change the dynamics of the interaction between the agent and the environment, but rather just to `observe' (at least, in the case of a classical interaction). To this end, all the maps of the tester are controlled unitary maps satisfying
\EQ{
U^T_k \ket{x}_{R_C} \otimes \ket{\psi}_{R_T} = \ket{x}_{R_C} \otimes U^{x}_k\ket{\psi} \label{tester}
} 
where $x \in \mathcal{S} \cup \mathcal{A},$ and $\{ U^{x}_k\}_x$ are arbitrary unitary maps acting on (the subsystems of) $R_T$, for all $k$.

\begin{figure}[!h]
\begin{equation}
\Qcircuit @C=0.7em @R=1.2em {
\lstick{R_{\tilde A}}&{/} \qw       & \qw                                 & \qw     & \multigate{1}{\mathcal{M}^A_1}& \qw  & \qw                                 & \qw                &\qw&\cdots   &  & \multigate{1}{\mathcal{M}^A_t}&\qw & \qw                                           &\qw &\qw       &\cdots    \\
\lstick{R_C \slash\, \substack{R_{I(A)}\\R_{I(E)}}}&{/} \qw          & \multigate{1}{\mathcal{M}^E_1}  & \ctrl{3} &\ghost{\mathcal{M}^A_1 }          & \ctrl{3}      & \multigate{1}{\mathcal{M}^E_2}  &  \ctrl{3} &\qw&\cdots  &   &\ghost{\mathcal{M}^A_t  }      & \ctrl{3}& \multigate{1}{\mathcal{M}^E_t}         &\ctrl{3}&\qw &\cdots   &  \\
\lstick{R_{\tilde E}}&{/} \qw            &\ghost{\mathcal{M}^E_1  }            &\qw      &  \qw                              & \qw   & \ghost{\mathcal{M}^E_2 }          & \qw            &\qw &\cdots      && \qw                             &  \qw & \ghost{\mathcal{M}^E_t  }                   & \qw&\qw & \cdots &\\
&&\-&\-&\-&\-&\-&-\ -\ -\ -\ -\ -\ -\ -\ -\ -\ -\ -\ -\ -\ -\ -\  -\ -\ -\ -\  -\  -\  -\  -\ -\ -\ -\ -\ -\  \\
\lstick{R_T}&{/} \qw & \qw         &\gate{U^T_2} &  \qw                                 &\gate{U^T_3}   & \qw         & \gate{U^T_4}    &\qw          & \cdots      && \qw &   \gate{U^{T}_{2t-1}}     & \qw &\gate{U^T_{2t}} &\qw & \cdots &
}\nonumber
\end{equation}
\caption{\label{fig2}
Tested agent-environment interaction for the communication/embodied model. Note that, in general, each map of the tester $U^T_k$ acts on a fresh subsystem of the register $R_T$, which is outside the control of the agent or the environment. The environments and agents $R_A,$ $R_E$ are subsystems of $R_{\tilde A},$ $R_{\tilde E}$, respectively, along with the purifying registers (possibly) needed for the unitary representation of the maps. The maps of the tester can be assumed be unitary. Each quantum ``wire'' corresponds to an arbitrary number of quantum systems (denoted with the ``$\slash$'' symbol on the wire).}
\end{figure}
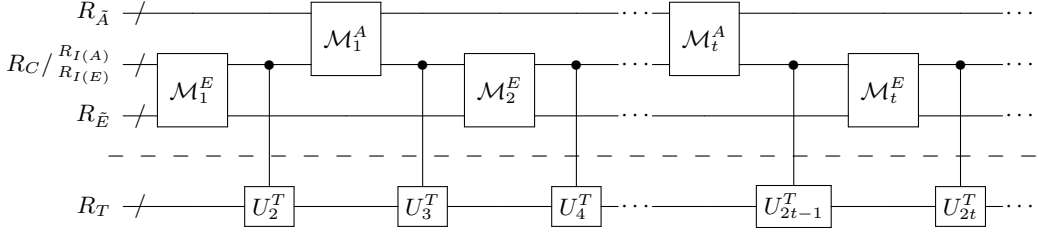

A \emph{classical tester} copies\footnote{In this work, by copy, we mean a unitary map which implements $\ket{x}\ket{\epsilon} \rightarrow \ket{x}\ket{x},$ where $\ket{x}$ is a percept or action basis state. If the duplicate is traced out, the realized map is just a classical basis measurement of the input state.} all the states of the $R_C$ register to its own.
A tested interaction between the agent and the environment, for the communications model, is illustrated in Fig. \ref{fig2}.

%\begin{figure}[!h]
%\label{fig2}
%\begin{equation}
%\Qcircuit @C=0.7em @R=1.2em {
%\lstick{R_A}&{/} \qw & \multigate{1}{U^A_1}  &\qw& \qw                                 & \qw     & \multigate{1}{U^A_2}& \qw  & \qw                                 & \qw                &\qw&\cdots   &  & \multigate{1}{U^A_t}&\qw & \qw                                           &\qw &\qw       &\cdots    \\
%\lstick{R_C}&{/} \qw &\ghost{U^A_1  }            &\ctrl{3}& \multigate{1}{U^E_1}  & \ctrl{3} &\ghost{U^A_2  }          & \ctrl{3}      & \multigate{1}{U^E_2}  &  \ctrl{3} &\qw&\cdots  &   &\ghost{U^A_t  }      & \ctrl{3}& \multigate{1}{U^E_t}         &\ctrl{3}&\qw &\cdots   &  \\
%\lstick{R_E}&{/} \qw & \qw                                &   \qw   &\ghost{U^E_1  }            &\qw      &  \qw                              & \qw   & \ghost{U^E_2 }          & \qw            &\qw &\cdots      && \qw                             &  \qw & \ghost{U^E_t  }                   & \qw&\qw & \cdots &\\
%&\\
%\lstick{R_T}&{/} \qw & \qw                                &\gate{U^T_1}& \qw         &\gate{U^T_2} &  \qw                                 &\gate{U^T_3}   & \qw         & \gate{U^T_4}    &\qw          & \cdots      && \qw &   \gate{U^{T}_{2t-1}}     & \qw &\gate{U^T_{2t}} &\qw & \cdots &
%}
%\end{equation}
%\caption{Tested agent-environment interaction for the communication model.}
%\end{figure}

The classical history is then recovered by the sequence of states of the register $R_T$, relative to the classical tester. A general quantum history is given by considering the state of the register $R_T$ without placing any (additional) restrictions on the maps of the tester, aside from the fact that we require they are of the `classically controlled' form given in Eq. (\ref{tester}). 
In the case of the embodied picture, there are two natural possibilities for the classical tester, standard and strong.
In the standard version, after the map of the environment (agent) is applied, the map of the tester follows, and it is a controlled map, controlled only by the interface of the environment (agent). Hence, the tester accesses the information of just one interface at a given time step.
In the strong version,  the states of both interface are accessed at each time step.  The standard variant of the classical tester in the embodied representation recovers histories of the same form as the classical tester in the communication model. In the strong variant, the states of each interface are, in general, copied twice. We will in this paper, unless stated otherwise, be using the standard tester for the embodied model. 
%Hence all the statements we will be making will hold for both the communication and the embodied model.

A few remarks are in order. In the case of stochastic classical agents (environments), the agent (environment) will, at each time step, output a particular action (percept) with some probability. In the quantum model, this will be represented by the agent outputting a convex mixture of action states, specified by the corresponding probabilities of the particular action. Thus, in the setting of a classical tester $T_c$, the state of the register $R_T$, at time-step $t$, can be expressed, in terms of the classical agent-environment interactions, as follows:
\EQ{
\rho^t_{T_c}(A,E) = \sum_{h_t \in \mathbf{H}_t} P(A \leftrightarrow_t E = h_t) \dm{h_t},\label{qt}
}

The above is exactly the classical history, defined in Section \ref{class-inter}, represented in the standard quantum formalism. 
The quantum history state $\rho^t_T(A,E)$ will, in general, attain significantly different forms for different testers, and we will refer to it as \emph{the quantum history between agent and the environment, relative to the tester $T$}. In the quantum interaction case, a figure of merit for learning will be a function of a quantum history of the interaction.

The presence of a classical tester changes nothing in the case of classical agents and environments. It is also not problematic for agents and environments which  only have a classical interaction, which is slightly more general:

\LE
\label{LE-equiv}
For any agent $A$ and environment $E$,
$A$ and $E$ have a classical interaction if and only if the (reduced) state of the three registers of $R_{A}R_C R_E$ is the same in the presence and absence of a classical tester.
\EL

Examples of such `internally quantum' agents which interact classically is, for instance, a standard (classical input-classical output) quantum computer, where the environment would be the users.
Such (internally only) quantum agents and environments which have a classical interaction cannot offer different behaviors, compared to classical agents interacting with classical environments, relative to any tester:
\LE
\label{LE-clas-int}
For any agent $A$ and environment $E$, which have classical interaction (when untested), there exists 
a classical agent $A^c$ and a classical environment $E^c$, such that $\rho^t_T(A,E) = \rho^t_T(A^c,E^c),$ for any tester $T,$ and any history length $t$.
\EL
Proofs of the two lemmas above are given in the Appendix (Section \ref{appendix}). Note that the lemma above also implies that no improvements of any figure of merit which is a function of the history alone can be achieved by utilizing quantum mechanics, if the interaction is classical.
Similarly, in the presence of a classical tester, quantum improvements are also not possible:

\LE
\label{LE-clasTest}
Let $A$ and $E$ be any agent and any environment over compatible percept/action spaces.
Then there exists a classical agent $A^c$ and a classical environment $E^c$, such that $\rho^t_{T_c}(A,E) = \rho^t_{T_c}(A^c,E^c),$ for the classical tester $T_c,$ and any history length $t$.
\EL
\proof
Adding an additional classical tester (instead of just one) still generates the same quantum history within the original classical tester. However, tracing out the register of the additional tester reduces the interaction of $A$ and $E$ to a classical interaction, as all non-classical terms (off-diagonal components in the states of $R_C$) are removed by the trace-out. But then by Lemma \ref{LE-clas-int}, the same quantum interaction generated by a classical tester can be achieved by a classical agent and a classical environment. Note that we cannot use the same argument for other testers - adding a second classical tester may change the quantum history generated by another type of tester. \qed
 
The results above should not be particularly surprising -- classical interactions simply lack the capacity for sufficiently subtle control to allow for any quantum effects (including speed-ups) almost by definition.
Thus we will have to consider other types of testers to achieve improvements.
 In this work, we will focus on the \emph{sporadic} tester, which allows for periods of untested, fully coherent interaction, followed by classically tested interaction. While this is still a restricted setting, maintaining the tester fully classical at periods will allow for a straightforward comparison between quantum and fully classical agents. 
 
 We note that, alternatively, the framework of quantum combs \cite{2008_Chiribella} could be used to model the environment-agent-tester interaction, and indeed, the generality of that approach may become beneficial in settings with multiple agents. We also note that the authors of the last reference and collaborators in \cite{2013_Chiribella} use the terms `quantum tester' and refer to `histories of classical communication', albeit in a cryptographic context, which are remotely related to the ideas we have presented. 
 
In the remainder of the paper, we will use the term \emph{fully classical agent} to refer to an agent which is classical, but also forces the interaction (for any environment) to be classical. Here, forcing implies that, within the model, the agent always dephases the register $R_C$ (equivalently, registers $R_{I(A)}R_{I(E)} $), by e.g. classical basis measurements, whose outcomes are discarded. 
Since, for the purposes of this work, we are interested in quantum enhancements of classical learning agents, in the next section we consider what kinds of quantum extensions classically specified environments in principle allow.

\subsubsection{The generic performance of a quantum-enhanced agent}
\label{generic-performance}

Suppose we are faced with a classical learning scenario, with a fully classical agent $A$ and an environment $E$, which is, a-priori unknown.
We would then like to asses the properties of the interaction of a quantum agent $A^q$, for the purposes of comparison,  with the \emph{same} environment $E,$ which can now be accessed via a quantum (not classical, in the sense of the definition we have in the previous section) interaction.
The question then is, in general, when can we consider two environments $E$ and $E'$ to be `the same'. There are a few natural answers.
The strongest notion of sameness would demand that two environments are equal, if they are specified by the same sequence of CPTP maps. A weaker notion of sameness is \emph{equivalence relative to the tester $T$}: $E$ and $E'$ are equal relative to $T$ if the quantum histories of $E$ and $E'$, relative to the tester $T$, are the same for any agent. If two environments satisfy the stronger notion of sameness, then they are equal relative to all testers. Note that all environments are equal relative to trivial testers, which apply the same map irrespective of the states in the communication register.
%A similar statement can be made in reverse.
However, since we are adopting the approach of extending classical learning scenarios to quantum for the purpose of comparison, we are interested in the following definition:
\DE
Two environments $E$ and $E'$ are the equal \emph{in the classical sense}, denoted $E=_c E'$ if they are the equal relative to the classical tester.
\ED
The above is equivalent to saying that $E$ and $E'$ are the same in the sense of realizing identical classical distributions over histories for any fully classical agent. The definitions above could also be relaxed to approximate equalities (within some distance) by relaxing the equalities on the quantum histories (using e.g. an approximate equality on states induced by the trace distance).

It is easy to see that the equality in the classical sense is an equivalence relation on environments.
For each environment $E$ we can then identify the classical equivalence class $\mathcal{E}_c(E) = \{E' | E =_c E' \}.$ All the elements of the class $\mathcal{E}_c(E)$ share the property that the classical maps they realize (in the sense of the classical definition of agent-environment interaction), in a classical interaction, are equal for all environments in the class.
This sequence of classical maps (i.e. this classical environment) we will call \emph{the classical specification} of the class $\mathcal{E}_c(E)$. Then we will also say an environment $E$ is \emph{only classically specified} if only its classical specification is known. 
Recall, in fully classical learning, classical specification is all there even is.
The next simple lemma states that if only the classical specification of an environment is known, no quantum enhancement can be generically guaranteed. 
\LE
\label{lem-extens}
Let $\mathcal{E}_c(E)$ be the classical equivalence class for some environment $E$. Then there exits a quantum environment $E^q \in \mathcal{E}_c(E)$ which prohibits any quantum improvement -- that is, any possible quantum history (relative to any tester), can be realized with a fully classical agent and this environment $E^q.$
\EL
\proof Take any environment $E' \in \mathcal{E}_c(E),$ and sandwich every CPTP map which specifies the environment $E'$ with a classical basis measurement of the register $R_C$ (equivalently, $R_{I(A)}R_{I(A)}$).
This is a new environment, $E^q,$ it is clearly in $\mathcal{E}_c(E)$, but it also forces a classical interaction. Then by Lemma \ref{LE-clas-int}, no quantum advantage is possible in this environment for any agent. \qed

The lemma above should be clear. With the permission of a bit of poetic license, it asserts that just putting on our ``quantum eyeglasses", that is, acknowledging that any real system is a  quantum system with quantum degrees of freedom, does not turn, for instance, a classical computer into a quantum computer. Even with fully coherent quantum input, most devices (or environments) will have decoherence processes which prevent any true quantum dynamics on any useful scale.
While this observation is straightforward, it is nonetheless relevant for our case. In what follows, we will begin by specifying an interaction between a classical environment and a quantum agent. Then, we will ask whether the quantum agent could do better, if the environment can be accessed as a quantum system, and the agent is free to exploit quantum coherence. Lemma \ref{lem-extens} then asserts that the answer may be a trivial no, unless further assumptions are made on \emph{how} the environment extends to a full quantum system\footnote{We acknowledge that, from a modern physics point of view, it would be more natural to consider this problem in reverse. Any physical system is fundamentally quantum, and one can consider classical limits of the quantum system, rather than `quantum extensions' of an otherwise classical systems. However, in the spirit of the mainstream approaches to artificial intelligence, systems, and task environments are usually assumed to be classical, both in the computational tradition and in robotics.
From this perspective, since we start from such classical problem, it makes sense to talk about quantum extensions, that is, quantum systems which are compatible with the given classical limit. 
}.

The question of what are useful quantum extensions of classically specified \emph{functions} is also vital in the case of quantum computation with the aid of a quantum oracle. Such a computation also falls within the scope of quantum agent-environment interaction, and we shall use results from quantum oracle models later.
We further investigate the relationship between (quantum) oracles, environments and learning in the remainder of the paper.

\section{Quantum oracles, agent-environment framework and learning}
\label{QAEI}

In this section we examine the relationship between oracular quantum computing models on one hand, and searching and agent-environment tasks and learning, on the other hand.

The two basic types of oracles \cite{2002_Kashefi} are characterized (relative to a a classical oracular boolean function $f$) by the following two expressions:
\EQ{
\ket{x}\ket{y} \stackrel{U^{(i)}_f}{\longrightarrow} \ket{x} \ket{f(x)\oplus y} \label{x-or}\\
\ket{x} \stackrel{U^{(ii)}_f}{\longrightarrow} (-1)^{f(x)} \ket{x} \label{z-or},
}
where $\oplus$ denotes addition in the underlying finite field (typically, mod 2 addition for the case of (qu)bits.).
A third type, which is more restrictive, yet still sufficient for some quantum Fourier transform-based algorithms (e.g. Shor's algorithm \cite{1994_Shor}, and other related algorithms) is specified by the following CPTP isometry:
\EQ{
\ket{x}\bra{x'} \stackrel{\mathcal{E}_f}{\longrightarrow} \ket{x}\bra{x'} \otimes \ket{f(x)}\bra{f(x')} \label{QFTor}.
}
The first oracle is strongest, that is, the other two can be reduced to it.
However, the first can be reduced to the second, if one additionally assumes quantum control can be added to the second oracle. The third oracle is the weakest, that is, the first and the second oracle cannot be reduced to it, given one oracle use.

In light of the previous section, we note that one can specify another ``quantum'' oracle, a CPTP map $\mathcal{F}_f$, which in every sense captures the classical function $f$ (is a valid quantum extension of $f$), yet is useless for any quantum algorithm.
 It is specified as follows:
 \EQ{
\mathcal{F}_f (\rho_{\textup{I}}) = Tr_{{\textup{I}}}\left[   U^{(i)}_f   \left( \rho_{\textup{I}} \otimes \dm{0}_{{\textup{II}}}  \right) (U^{(i)}_f)^{\dagger}   \right]. \label{bad-oracle}
 }
The map $\mathcal{F}_f $ given a pure state input $\ket{x}$, returns the result $\ket{f(x)},$ but given any superposition of inputs only returns the (classical) convex mixture of classical results. 
The utility of various types of oracles has been thoroughly investigated and it is well-established that the particular format of the oracle can greatly influence the query complexity of oracular algorithms. For instance, an exponential separation for the set comparison problem is reported in \cite{2002_Aaronson}, based on a non-standard oracle defined in \cite{2002_Kashefi}. 

In the remainder of the section we consider a similar type of a question, and aim at characterizing the environments which can be reduced to one of the first two oracular types with the goal of using Grover-like (or, more generally, amplitude amplification) approaches which we later show can facilitate faster learning.
%\red{Importantly we will hint at the potential of the third type of oracle, which will be used in the second paper.}
However, there are three significant obvious distinctions between agent-environment interactions and oracles which we will have to take into account.

First, standard oracles are one-step maps - environments in general produce a relevant output after sequences of interactions. Second, standard oracles have no memory - environments, in general, have memory. Third, the register that the standard oracular map is applied to is a part of the system we have full control over - environments are embodied, and in general, the agent has no access to the internal registers of the environment.
These three distinction severely restrict the applicability of oracle-based results: in particular, the last two imply that, in general, the tracing out of the environment breaks any superposition the agent may have used, which kills any speed-up, regardless of the tester. However, one may still hope that at least some quantum extensions of classically specified environments may allow for non-trivial quantum interactions.

As it turns out, even the most amenable quantum extensions of classically specified environments make poor oracles in general.
%Under the stringent constraints of the quantum agent-environment interaction model, even in the best but non-trivial scenarios, one can achieve the oracles of the type given in Eq. (\ref{QFTor}), the utility of which is not straightforward in learning contexts. This we elaborate to further detail in the Appendix, section \ref{sec-envOr}. For now, we are interested in methods which will aid us in treating the environment as the strongest (most useful) type of a quantum oracle.

In the following, we consider what it would mean, for the purposes of this paper,  to have environments which make \emph{good oracles}, and show how this can be utilized in learning. This will be immediately applicable to constructed, or artificial environments, where the agent can influence the design of the environment to match its purposes. A particular powerful method of learning, model-based learning, which we discuss later, will be an example where (internally) constructed environments play a central role. 

\subsection{Oraculazing constructed environments}
\label{good-oracles}
One of the characterizing features of learning  task environments is the reward function $R$, which specifies which sequences of interactions are rewarded. On an intuitive level, the task of any learning agent is to, over time, start contributing toward histories which do yield rewards\footnote{We say ``contributing toward histories'' as the reward may depend on the percepts issued by the environment, and this may be stochastic, and beyond the powers of the agent to perfectly control.}.
We then wish to view the environment as an oracle, instantiating this reward function. This oracle is then queried through sequences of actions of the agent, in an attempt to find the best sequences of actions. 

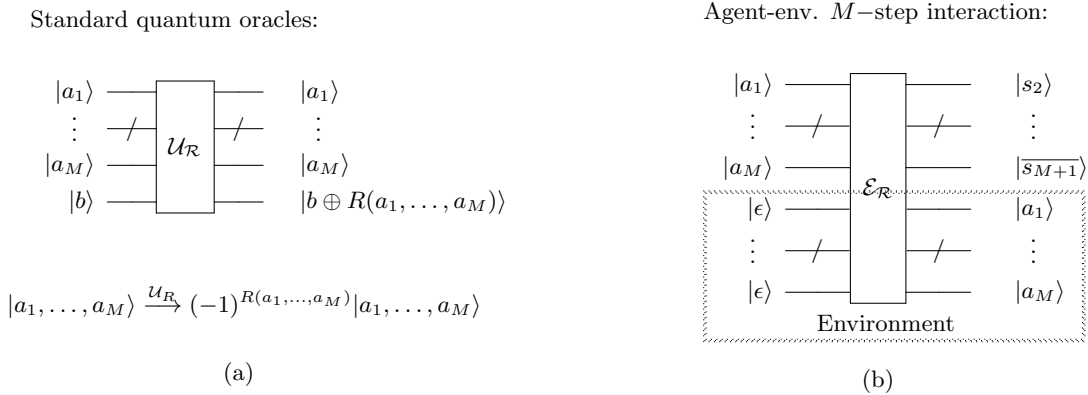
\begin{figure}
%\begin{flushleft}
\begin{minipage}{.6\textwidth}
%\begin{flushleft}
\hspace{-5cm}Standard quantum oracles:
\\
\vspace{0.5cm}
\hspace{-4.5cm}\mbox{
\Qcircuit @C=1em @R=0.6em {
\lstick{\ket{a_1}}&\qw& \multigate{3}{\mathcal{U_R}} & \qw &\qw&\rstick{\ket{a_1}} \\
    \lstick{ \raisebox{0.6em}{ \vdots} \hspace{0.7em}   }         &{/}  \qw &\ghost{\mathcal{U_R}}& {/} \qw&\qw  & \rstick{\hspace{0.3em}\raisebox{0.6em}{ \vdots}    }  \\
\lstick{\ket{a_M}}&\qw& \ghost{\mathcal{U_R}} & \qw &\qw&\rstick{\ket{a_M}}\\
\lstick{\ket{b}}              &\qw & \ghost{\mathcal{U_R}} & \qw &\qw&\rstick{\ket{b \oplus R(a_1, \ldots, a_M)} }\\
\\
\\
\\
\\
\\
&&&&\hspace{-0.5cm}\ket{a_1, \ldots, a_M} \stackrel{\mathcal{U}_R}{\longrightarrow} (-1)^{R(a_1, \ldots, a_M)}\ket{a_1, \ldots, a_M}\\
\\
\\
\\
\\
&&&\textup{(a)}\\
}
}
%\end{flushleft}
\end{minipage}%
%\end{flushleft}%
\begin{minipage}{.7\textwidth}
\hspace{-6cm}Agent-env. $M-$step interaction:
\\
\vspace{0.5cm}
\begin{centering}
\Qcircuit @C=1.3em @R=0.8em {
&&\lstick{\ket{a_1}}&\qw& \multigate{5}{\mathcal{E_R}} & \qw &\qw&\rstick{\ket{s_2}} &\\
 & &  \lstick{ \raisebox{0.6em}{ \vdots} \hspace{0.7em}   }         &{/}  \qw &\ghost{\mathcal{U_R}}& {/} \qw&\qw  & \rstick{\hspace{0.3em}\raisebox{0.6em}{ \vdots}    } &  \\
&&\lstick{\ket{a_M}}      &\qw& \ghost{\mathcal{E_R}} & \qw &\qw&\rstick{\ket{\overline{s_{M+1}}}} &\\
&&\lstick{\ket{\epsilon}} &  \qw  & \ghost{\mathcal{U_R}}  & \qw &\qw&\rstick{\ket{a_1}} & &  \\
& &\lstick{ \raisebox{0.6em}{ \vdots} \hspace{0.7em}   }         &{/}  \qw &\ghost{\mathcal{U_R}}& {/} \qw&\qw  & \rstick{\hspace{0.3em}\raisebox{0.6em}{ \vdots}    } & \\
 &&\lstick{\ket{\epsilon}} & \qw  & \ghost{\mathcal{E_R}}  & \qw &\qw&\rstick{\ket{a_M}}&&\push{} \gategroup{4}{1}{7}{10}{1.5em}{..}\\
 & &  &   & &\hspace{-1.3cm}\textup{Environment }& && &\push{} \gategroup{4}{1}{7}{10}{1.4em}{..} \gategroup{4}{1}{7}{10}{1.3em}{..}\\
 \\
 \\
&&&&\textup{(b)} \\
 }
\end{centering}
\end{minipage}

\caption{\label{fig4}
Environments as oracles. (a) The two standard quantum oracles defined for the underlying reward function $R$ of the environment. (b) A possible representation of the actual process realized by agent-environment interactions: the agent receives the percepts issued by the environment, but since the reward (in this case, rewarded or non rewarded subspace of~$\overline{s_{M+1}}$) depends on all the actions of the agent, the environment must have retained the actions (as illustrated) or copied the actions of the agent to its register. In the latter case, any superposition state of actions would be broken, from the perspective of the agent, that is, when the environment is traced out.}
\end{figure}

Oracles are typically deterministic. As our first constraint, we shall thus consider fully deterministic environments, which also implies that any sequence of actions of the agent uniquely specifies the subsequent percept of the environment. For simplicity we will also consider binary rewards only, but this is a less problematic assumption as we comment later.  
In this case, there exists a (set of) shortest deterministic sequence(s) of actions $(a^r_1, \ldots, a^r_{M})$ which the agent can perform, starting from the first time-step, and which will yield a reward. 
The corresponding quantum oracles, which would allow to perform a Grover's algorithm \cite{1996_Grover} based search for the right actions, are presented in Fig. \ref{fig4}~(a). The top oracle in Fig. \ref{fig4}~(a) we will refer to as the `bit flip' oracle, and the lower as the `phase flip' oracle, for convenience. 
Unfortunately, the environment will have to store the sequence of actions (while returning the percepts), illustrated in Fig. \ref{fig4}~(b), or at least copy them to its internal memory, as the reward may depend on $M-$long sequence of actions. The latter option, copying, does not remedy the problem as the tracing out of the environment breaks any superposition of action states the agent might have output.

 However, in the case the environment is constructed, and in the case we really do not require the percepts of the environment, we may assume the environment is given in an oracular form of, say, the phase flip oracle.

As our next observation, note that for quantum oracles to be useful, one typically requires more than one access to it. Notable exceptions to this are the Deutsch-Jozsa \cite{1992_Deutsch} or related algorithms, where certain promise problems can be solved with just one access. If our goal is to asses the proper sequence of actions, in general we require multiple accesses. 

Furthermore, environmental outputs may depend on entire histories. In particular, this means that although a given sequence may have been rewarded, it no longer needs to be if the same sequence is repeated. This is analogous to having the oracle dynamically change through sequential use, moreover, the length of the input to the oracle may change as well. The situation is further convoluted in the case the agent attempts to use superpositions of actions as inputs, as the corresponding oracles will then, in general, be applied in superpositions as well.
Luckily, many relevant task environments, specifically in the context of reinforcement learning, are assumed to be periodic, or epochal.
For instance it is often the case that, once the reward is given, the environment `resets' itself to the initial condition. This is equivalent to having the reward map depend only on the most recent history, only back to the time step of the last rewarded action.
Examples of such task environments are maze-like problems, where an agent is supposed to find a path from a starting position to a goal. Then for the agent to learn in such a setting, the same maze is traversed anew, after the goal has been reached. We will call such an episode \emph{a game} or \emph{a learning epoch}.
These seemingly restricted environments nonetheless contain very interesting real-world examples. For instance, in the case of common two-player board games (e.g. go, chess) playing against a deterministic opponent can be rephrased in such a manner.

For our immediate goal, we shall add one more assumption: namely, that the environment allows only \emph{single-win fixed-time games}.
\emph{Fixed-time} implies that, regardless whether the agent won or not, a single game lasts $M=M_{max}$ steps. \emph{Single-win} implies that (at most) one percept is rewarding in one game.
The single-win assumption is important as it eliminates a parity problem which would otherwise arise in the using of phase flip oracles - being rewarded twice would just equate not being rewarded. The fixed-time assumption guarantees that the actions of the agent do not leak into the next game, \emph{e.g.} in the case of a fast win.
We shall comment on how these assumptions can be removed later.

For any single-win fixed-time deterministic environment $E,$ we can thus construct $E^{q}_{oracle}$, relative to the reward function $R$ of $E$, as given in Fig. \ref{fig4} (a). In the following we will refer to the phase flip oracle, unless stated otherwise.
We remind the reader that $E^{q}_{oracle}$ is not a proper quantum extension of $E$, that is, not in the set $\mathcal{E}_c(E)$.

Given a single-win fixed-time deterministic environment $E$, with $E^q_{control}$ we denote a quantum environment, where, via an additional control mechanism, the agent can choose whether $E^q_{control}$ behaves like a proper quantum extension of $E$ or like the corresponding oracularized variant $E^{q}_{oracle}$. We will call such environments \emph{controllable environments}.

\section{Faster learning for artificial environments: an example with a maze}
\label{main-section}

{In the next two sections we} show that having controllable environments can help in learning.
{We begin by first clarifying} \emph{what is that which can be learned} given the model we have presented. The static representation of $E^{q}_{oracle}$, of a given environment $E$ suggests that the only thing to be done is to obtain an action sequence which leads to a reward.
However, that is typically only an initial ingredient of an overall learning goal.

Recall that real task environments do change in time, in particular, the winning sequence may change over the course of many epochs or games (e.g. the opponent changes its strategy). The change, however, usually has an underlying structure to it. The task of the agent is then to utilize the experience it obtained in the previous games to its advantage in the future - that is, to learn the relevant \emph{behavioral patterns} appropriate for the given task environment.

It will be illustrative to give one concrete and comprehensive example - a maze-like game.

Consider a connected directed graph, with regular out-degree $n$, over the set of $N$ vertices. Each vertex is labeled from a set of labels $L$.
Two particular vertices are labeled $Start$ and $Finish$.
Next, imagine an agent, who is initially positioned at the vertex $Start$. It can perform $n$ different moves (from the $n$-element set $D$) to one of the $n$ neighbors of the current vertex, and its task is to walk to the labeled $Finish$. We assume which move leads to which neighbor is specified for the graph.  The minimal path length from $Start$ to $Finish$ is $M$, and after exactly $M_{max} \geq M$ steps, the agent is always teleported to the $Start$ vertex, and is rewarded if at any point it actually encountered the $Finish$ vertex.

\begin{figure}[!h]
\centering
%%
%\begin{minipage}{.5\textwidth}
%\hspace{-0cm}\includegraphics[width=0.9\textwidth, clip=true, trim = 10 300 550 10]{Maze1.pdf}%
%\end{minipage}%
%%\begin{flushright}%
%\begin{minipage}{.5\textwidth}%
%\includegraphics[width=0.9\textwidth, clip=true, trim = 10 300 550 10]{Maze2.pdf}
%\includegraphics[width=0.9\textwidth, clip=true, trim = 10 300 550 10]{Maze3.pdf}
%\end{minipage}
%%\end{flushright}
%, clip=true, trim = 10 300 550 10
%, clip=true, trim = 10 300 550 10]
\begin{minipage}[b]{0.47\textwidth}
\fbox{\includegraphics[width=\textwidth, clip=true, trim = 20 60 145 60]{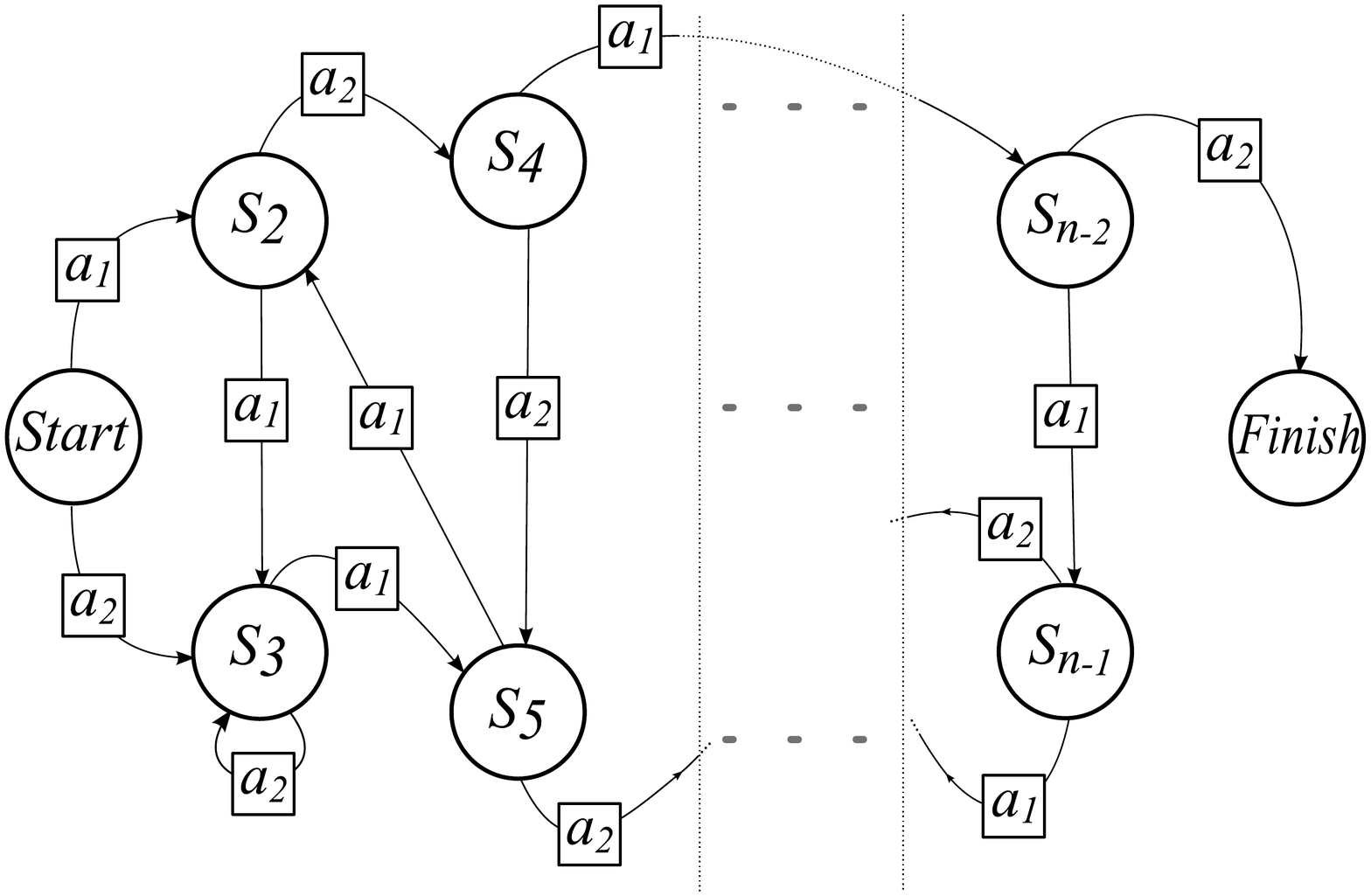}}
\vspace{0.25cm}

(a) 
\end{minipage}
\hspace{0.4cm}
\begin{minipage}[b]{0.47\textwidth}
\fbox{\includegraphics[width=\textwidth, clip=true, trim =  20 60 145 60]{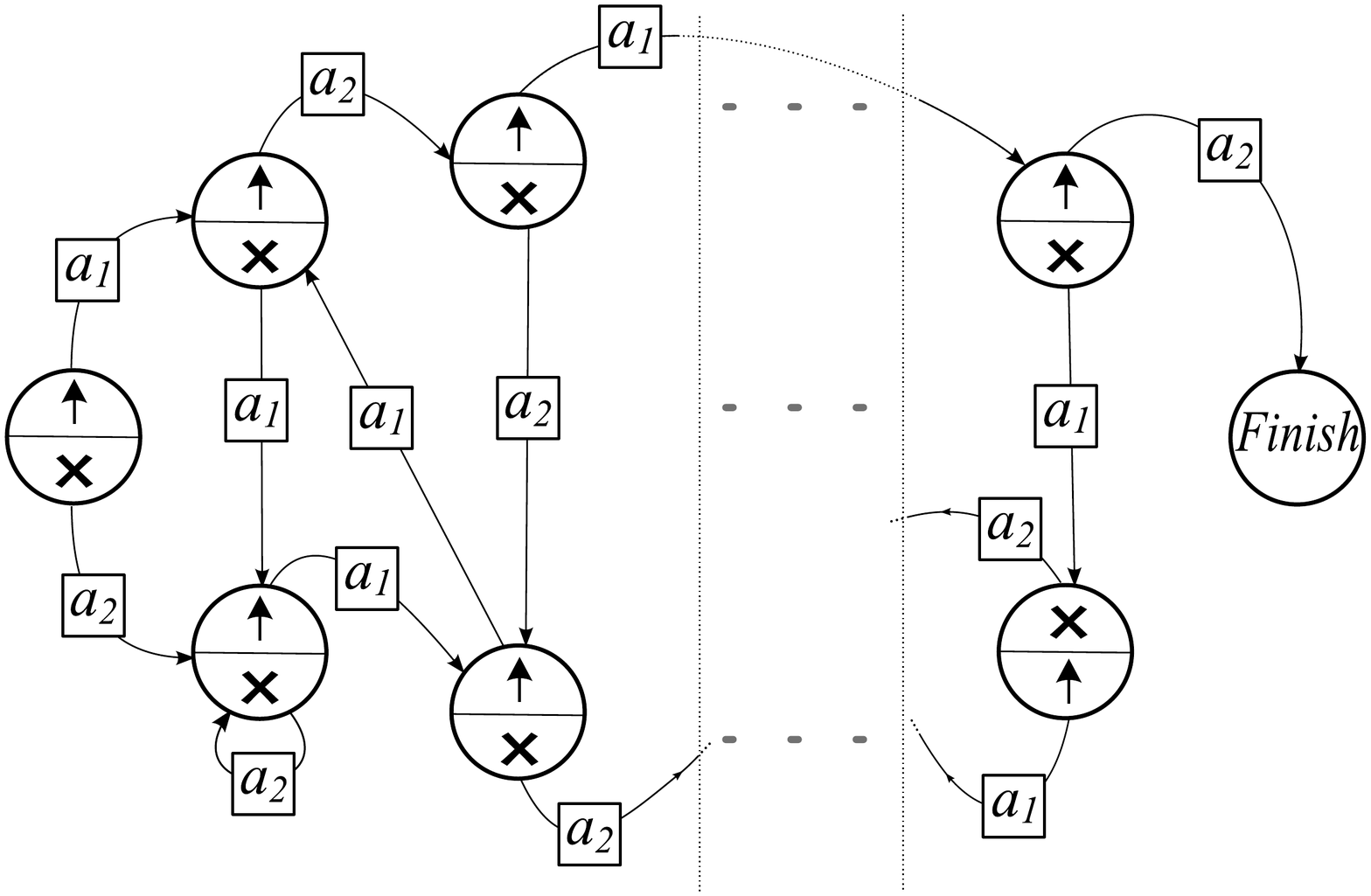}}
\vspace{0.25cm}

(b) 
\end{minipage}
%\fbox{\includegraphics[width=0.7\textwidth, clip=true, trim = 10 300 550 10]{Maze3.pdf}}

\caption{\label{fig-mazes}
Figure (a) depicts the abstract problem setting for an involved underlying graph. The agent can choose between two actions, indicated on the directed edges. The labels of the vertices indicate the percepts provided by the environment. In figure (b) we add the semantics of the problem - the information about the correct moves for the agent is actually provided by the environment: the percepts $(\uparrow, \times)$ and $(\times, \uparrow)$  inform the agent that the first (say "left") or the second (say "right") action leads to the $Finish$ vertex, respectively. In the figure this has been depicted by the relative positioning of $\times$ and $\uparrow$ within the vertex, for illustration purposes. For instance, the percepts corresponding to the first three percepts ($Start$, $S_2$, $S_3$ in Fig. (a)\ ) would be $(\uparrow, \times), (\times,\uparrow)$ and $(\uparrow, \times),$ respectively, where the position in the ordered pair specifies the action ($a_1$ for the first, $a_2$ for the second position).
The agent is expected to learn the behavioral pattern ``follow the arrows'', using which it can easily resolve any subsequent maze with the same percept structure, without having to search. Nonetheless, for the first game, the agent has no other recourse but to blindly search for the $Finish$ vertex. 
}
\end{figure}

The scenario as described is a deterministic, single-win fixed-time scenario we wish to consider. {This task environment, for the case $n=2,$ is illustrated in Fig. \ref{fig-mazes} a), with the understanding that only a fixed number of steps are allowed before the walker is re-set to the initial vertex $Start$.}
{Even at this abstract level, it is still clear that learning agents can still find the target vertex, and, even better, over iterations of the game, eventually learn to follow the optimal path. As an example, consider a simple agent, that simply walks randomly, until the reward is given. Its only update rule stipulates that once the reward has been given, the agent will, from that point on, deterministically re-produce the last sequence of $M$ steps.
For illustrative purposes we will focus on the case $n=2$.
For this agent we can explicitly give the distribution over histories $A\leftrightarrow_M E$ of the first $M$ steps:}
\EQ{
P \left(A\leftrightarrow_M E  = h_{a_1, \ldots, a_M} \right) = 2^{-M}, \textup{with} \label{distr}\\
 h_{a_1, \ldots, a_M} =\left(Start, a_1, s\left(a_1\right), a_2, s\left(a_2,a_1\right)\ldots a_M, s\left(a_M, \ldots a_1\right) \right)\nonumber
}
{where $s(a_1, \ldots, a_k)$ specifies the vertex (or ``room'' if we understand the graph as a map of ``rooms'' - vertices, connected by ``doors'' - actions) we end up in, given the sequence of moves $a_1, \ldots, a_k$. Note that we are, for the time being, considering this a classical environment, so a notion of history is well defined without a tester.

Since the environment is deterministic, the history is fully determined just by the actions of the agent.
 Equivalently, if we view the agent-environment interaction as a quantum exchange relative to the classical tester, we recover Eq. (\ref{qt}) for the (uniform) distribution $A \leftrightarrow_M E$ above.}
{The probability the agent finds the $Finish$ vertex is simply given by the sum of the probabilities of all histories which have $Finish$ as the final percept/vertex.
In the easiest-to-illustrate case when $M_{max} = M$, our simple agent shall then, effectively, sample from the distribution $A\leftrightarrow_M E$ (by trying out random paths), until it does encounter vertex $Finish.$ Following this the distribution over histories becomes a Kronecker-delta distribution, with the entirety of mass on the one history the agent is repeating.
In this simple example, ``learning" really did reduce to ``finding".
}

Now, we shall add a little more context. {Suppose that} the label set, that is, the percept set of the agent, contains {$2$} elements:
% \EQ{L=\{(\uparrow, \times, \ldots \times),(\times, \uparrow~, \times, \ldots \times), \ldots, ( \times, \ldots \times,\uparrow) \}.} 
\EQ{\mathcal{S} = L=\{(\uparrow, \times),(\times, \uparrow) \}.}

Unknown to the agent, the labels of the vertices specify whether the first, the second, or  the $k^{th}$ action (relative to some order) leads closer to the finish vertex.
This scenario is illustrated in Fig.~\ref{fig-mazes}~b).

Initially, the percepts mean nothing to the agent. 
However, an intelligent agent, having traversed the same graph many times and in perhaps different ways, will start to recognize the logic. 

{Suppose further that} after a certain number of games (the traversing of the graph), the underlying graph changes, but not the semantics of the percepts.

Our simple learning agent will have no useful knowledge to take going from such one labeled maze to the next. 
 And indeed, any agent which relies only on exhaustive search, constitutes a bad learning agent, in this context.
A good learning agent will, hopefully, have learned a valuable lesson -- that it should follow the signs.
In the example provided, we have given a strong structure to the underlying problem -- a one-to-one correspondence between percepts and the required subsequent actions. In general, the percept structure may be, to a lesser or greater degree, informative as to what the correct sequence of moves should be.
%The learning models we consider are positive-reward learning models (and most, including PS are such). For such models, we can define agent-
{However, both settings,  abstract percepts\!~-~\!no underlying structure\!~-~\!a fixed environment, and the setting where the percepts label the correct sequence of actions have a crucial point in common.}
Even in the case that the environment provides strong `hints' towards the correct sequence of actions, initially, a learning agent must first undergo a phase of exploration, before exploitation of the accumulated experience can commence. The finding the correct balance the two phases is a central topic in standard reinforcement learning~\cite{SuttonBarto98}.

In the example we have shown, it is clear that, essentially, \emph{there is nothing useful the agent could learn} before the $Finish$ vertex was hit for the first time\footnote{One could argue that already the fact that certain sequences of moves are \emph{not} rewarded is useful information, however, in all generality, the agent need not even know that there \emph{is} a rewarding sequence at all.}. 
{This establishes a lower bound on the expected number of steps which must be taken by any agent, before any learning could commence, and it exists in the classical and quantum case alike. The difference is that, in the quantum case, given a controllable quantum environment, we will show that this expected time for the first hit of the target vertex may be generically quadratically improved\footnote{Moreover, without imposing additional structure on the task environment, better than quadratic improvements are unlikely to be possible, by the same arguments of the optimality of the Grover's search algorithm \cite{1998_Boyer, 1999_Zalka}.}.}
{The improvement in the expected times of a first rewarded event, however, does not alone constitute an improvement in learning.
}

{Nonetheless, improving exploration times can establish a starting point for constructing better learners (more precisely, \emph{improved learners}) in some settings.
}
{To understand how this works, consider two identical (fully classical) agents which are solving the same maze-like problem, and one, by chance alone, chooses good paths initially, whereas the other chooses long paths. The luckier agent will accumulate the relevant data sooner. Consequently, the lucky agent may be expected to perform better in the given environment over longer time-scales, even though the underlying learning model is the same.
In what follows we will show how to combine (essentially) any learning model, with the capacity to perform the exploration phase by utilizing the aforementioned quantum advantage. 
}

\section{Luck-favoring settings}
%\vdc{tweaked title as only pair learning model - environment can be luck favoring}
\label{lfa}
We begin by formalizing the scenarios where the lucky agents do learn faster. In the following, with $A(h_t)$ we denote the agent $A$ who has undergone the history $h_t$, that is, with the internal configuration, \emph{i.e.} the memory of the agent modified accordingly. Note that $A(h_t)$ is a learning agent with the same percept-action structure as $A$.
Then we have the following definition.
%\begin{samepage}
\DE
Let $A$ be a learning model/agent and $E$ a legitimate (with matching percept-action structure) environment of A.

Let $Rate(\cdot)$ denote a learning-related figure of merit, defined on histories and extended to distributions over histories by convex-linearity (e.g. the average reward of a history per time-step).

Then we say that the pair $(A, E)$ is \emph{monotonically luck-favoring} for histories $h^{E}_t$ and $\tilde{h}^{E}_t$ relative to the merit function $Rate(\cdot)$ if 
\EQ{
Rate(h^E_n) \geq Rate(\tilde{h}^E_t) \Rightarrow Rate(A(h^E_t) \leftrightarrow E) \geq Rate(A(\tilde{h}^E_t) \leftrightarrow E), \label{main-eq-luck}}

where $h^{E}_t$ and $\tilde{h}^{E}_t$ denote two (classical) histories of length $t$ that could have been generated by an interaction of $A$ with $E$, thus:
 \EQ{P(A\leftrightarrow_{t} E = h^{E}_t) \neq 0\ & \textup{and}\nonumber\\ P(A\leftrightarrow_{t} E = \tilde{h}^{E}_t) \neq 0.& \nonumber }
If Eq. (\ref{main-eq-luck}) holds for any two histories, then we say $(A,E)$ is monotonically luck favoring for all histories.

More specifically, we may be interested in the behavior for specified numbers of interactions $t,t'$. Then we say that
$(A, E)$ is \emph{monotonically luck-favoring} for the merit function $Rate(\cdot),$ with an n-step preparation $(h^{E}_t, \tilde{h}^{E}_t)$, followed by $t'$ step evaluation if
\EQ{
Rate(h^E_t) \geq Rate(\tilde{h}^E_t) \Rightarrow Rate(A(h^E_{t}) \leftrightarrow_{t'} E) \geq Rate(A(\tilde{h}^E_t) \leftrightarrow_{t'} E) \label{main-eq-luck2}
}
%\blue{We will say  the pair $(A, \mathcal{S}_E)$ is \emph{strongly} monotonically luck-favoring if the implication above holds for histories of differing lengths $h^E_n$ and $\tilde{h}^E_{n'}$ with $n \not= n'$.}
\ED
%\end{samepage}
The definition above captures a relatively broad scope of what ``luck-favoring'' may mean and can be elucidated via a simple example. 
The statement of Eq. (\ref{main-eq-luck}) can be rephrased as follows. Suppose you have two instances of the same agent, $A$ and $A'$, both interacting with (two instances of) the same environment $E$, for some number of rounds $t$. Also suppose that $A,$ (by chance alone, since the agents are otherwise identical) has been performing better up to time-step $t$ relative to the figure of merit $Rate$. Now, if this alone implies that  (the lucky agent) $A$ will, from that point on, be performing better than the (unlucky agent) $A'$ \emph{on average}, then $A$ and $E$ are luck favoring for $Rate$ and all histories.

It is important to note that the agents $A$ and $A'$ have \emph{identical} underlying characteristic maps - with this in mind it should be clear that, provided the agent can learn at all in a given environment, it can only benefit from a bit of luck.

%\blue{The notion of strong luck favoring is not as easily guaranteed. Consider a lucky maze agent, which has found one winning path, and an agent which has, by sheer misfortune, tried out all the paths except for the winning ones.
%If the algorithm makes any sense, the latter agent will in fact learn faster, from that point on.
%}

For instance, the maze-like environments we have described will be luck-favoring for most reasonable learning models, most figures of merit and most histories - this is easiest to demonstrate on mazes where there is a unique path of length $M$ leading to $Finish$, whereas all other paths take exponentially many more steps, in the maze size.
{Now, if we consider again Eq. (\ref{distr}) specifying the history of the random agent. In the case of a unique winning path, the probability the agent encounters the $Finish$ vertex is $2^{-M}.$ The situation is worse in the case that the maze has more than two actions (``doors") $n>2$ - for this case, we obtain $n^{-M}$.
 Being unlucky (or rather, failing to be \emph{exceptionally} lucky) in this situation may imply the path is, effectively, never found.}

To present our result regarding the speed-up in learning in a clean form, we shall place additional assumptions on the environment  $E$ aside from it being deterministic, single-win and fixed-time. Additionally, we will assume that there is only one winning action path of the length $M,$ where $M$ is also the allotted fixed time (in this case, there is only one winning history of length $M$).
{Recall that, in the case of the maze environment we have described, this implies that the agent traversing the maze is always returned to the $Start$ vertex after exactly $M$ steps.}

 Let $n$ be the size of the action space, thus $n = |\mathcal{A}|.$
Then, the classical agent will require, on average, $O(n^M \times M)$ interaction steps with the environment, before encountering the winning path (note that each `testing' of a particular sequence of action costs $M$ interaction steps).
The quantum agent, given access to the oracular instantiation $E^q_{oracle},$ can achieve the same in expected time $O(\sqrt{n^M} \times M)$, using the standard Grover's algorithm \cite{1996_Grover,1998_Boyer}.
This constitutes a quadratic improvement in the exploration phase of learning, and what remains to be seen is how to embed this into the complete learning package.

Both the classical and the quantum agent we will now construct are situated in the same controllable environment, namely, $E^q_{control}$. The classical agent $A$ has nothing to gain from quantum oracular access\footnote{Note that since we are using the phase flip oracle and since the fully classical agent dephases the responses (\emph{i. e.} measures them in the classical basis), no information about the reward would come to the agent.} and its access to this environment is only via its classical instantiation $E$.

For the classical agent $A$ (and its underlying learning model) we will next define a corresponding quantum agent $A^q$. Following the precise specification, we will briefly comment on the basic ideas behind the construction.

Since $A$ is fixed and known, we will assume $A^q$ has black-box access to (a simulation of) the agent $A$. In particular, $A^q$ can, internally, feed the simulation of $A$ with any sequence of percepts, and observe the output actions. Moreover, it can always reset the simulation to the initial state as defined for the agent $A$.
Since we are constructing $A^q$ given a classical agent $A,$ we in principle have access to every aspect of $A$ (its program, realization and specification of each characteristic map), but for our purposes, black-box access, and the capacity to reset will suffice.
As a technical assumption, we will assume that the agent $A$ has a non-zero probability of hitting the rewarding sequence of actions, starting from its initial configuration. 
We give a formal specification of the quantum agent $A^q$ next, followed by an explanation of the purpose of each of the steps.

\begin{enumerate}
\item \label{prep1} For the first $t' = k \times\sqrt{n^M} \times{M} $ time steps, $A^q$ engages in a Grover-type search for the awarded sequence of actions, interacting with $E^q_{oracle}$. Recall that each access to the oracle incurs $M$ interaction steps, thus we total $ k \times\sqrt{n^M}$ oracular queries, where $k$ is an integer we specify later.
The agent $A^q$ succeeds in finding the winning sequence $(a_1, \ldots, a_{M}),$ except with probability in $O(\exp(-k))$\footnote{Recall, Grover's algorithm may fail to produce the target element, but this occurs with probability less than 1/2. Iterating the algorithm $k$ times ensures that a failure can occur at most with an exponentially decaying probability in $k$.}, since the fraction of winning versus the total number of sequences is $n^{-M}$.
\item \label{prep2} For the next $M$ time-steps, the agent $A^q$ engages the proper quantum extension $E^q$ of $E$, outputs the (classical) actions $a_1, \ldots, a_{M},$ sequentially and collects the unique corresponding outputs $s_2, \ldots s_{M+1}$ from the environment (by convention, we set the first percept of the environment to be the empty percept $\epsilon$). The entire rewarding history is \EQ{h_{win}= (s_1, a_1, s_2, \ldots, a_{M}, s_{M+1}).} This step is necessary as the oracular access, by construction, does not provide the perceptual responses of the environment.

\item \label{training} Between the time steps $t = t'+M$ and  $t = t'+M+1$, $A^q$ `trains' a simulation of $A$ internally:
It runs a simulated interaction with $A,$ by giving percepts $s_1, \ldots, s_{M+1}$. It aborts and restarts the procedure (with a reset of the simulation of the agent $A$)  until $A$ responds with $(a_1, \ldots, a_{M})$. 
By the technical assumption we mentioned earlier, the expected time of this event is finite.  
The training procedure itself, for the $M$ time steps, is repeated sequentially, until the same winning sequence of actions of the simulated agent $A$ is produced $1+k \sqrt{n^M}$ times, again contiguously. 
Since one sequence can be attained in finite time, so can any finite repetition of the sequence.
This technicality we further explain later.

During this time the agent $A^q$ does not communicate to the environment, and uses up no interaction rounds.
\item \label{trained} Internally, $A^q$ has a simulation of the agent $A(h_{tot})$, with
\EQ{
h_{tot}=\underbrace{h_{win} \circ \cdots\circ h_{win}}_{(1+k\times\sqrt{n^M})\ times}
}
and $\circ$ denotes the (string-wise) concatenation of histories. From this point on, $A^q$ simply forwards the percepts and rewards between the simulation and the environment. 
\end{enumerate}

To talk about learning properties of the defined quantum agent we need to specify the tester.
To optimize our result, we select the sporadic classical tester $T_S$ which is defined as follows:

For the first $t = k \times\sqrt{n^M}\times{M}+M$ time-steps, with $k \in \mathbbmss{N},$ the sporadic tester allows for completely untested interaction. After the $t$ steps, the tester $T_S$ behaves as the classical tester.

\begin{figure}
\centering

\fbox{\includegraphics[width=0.8\textwidth]{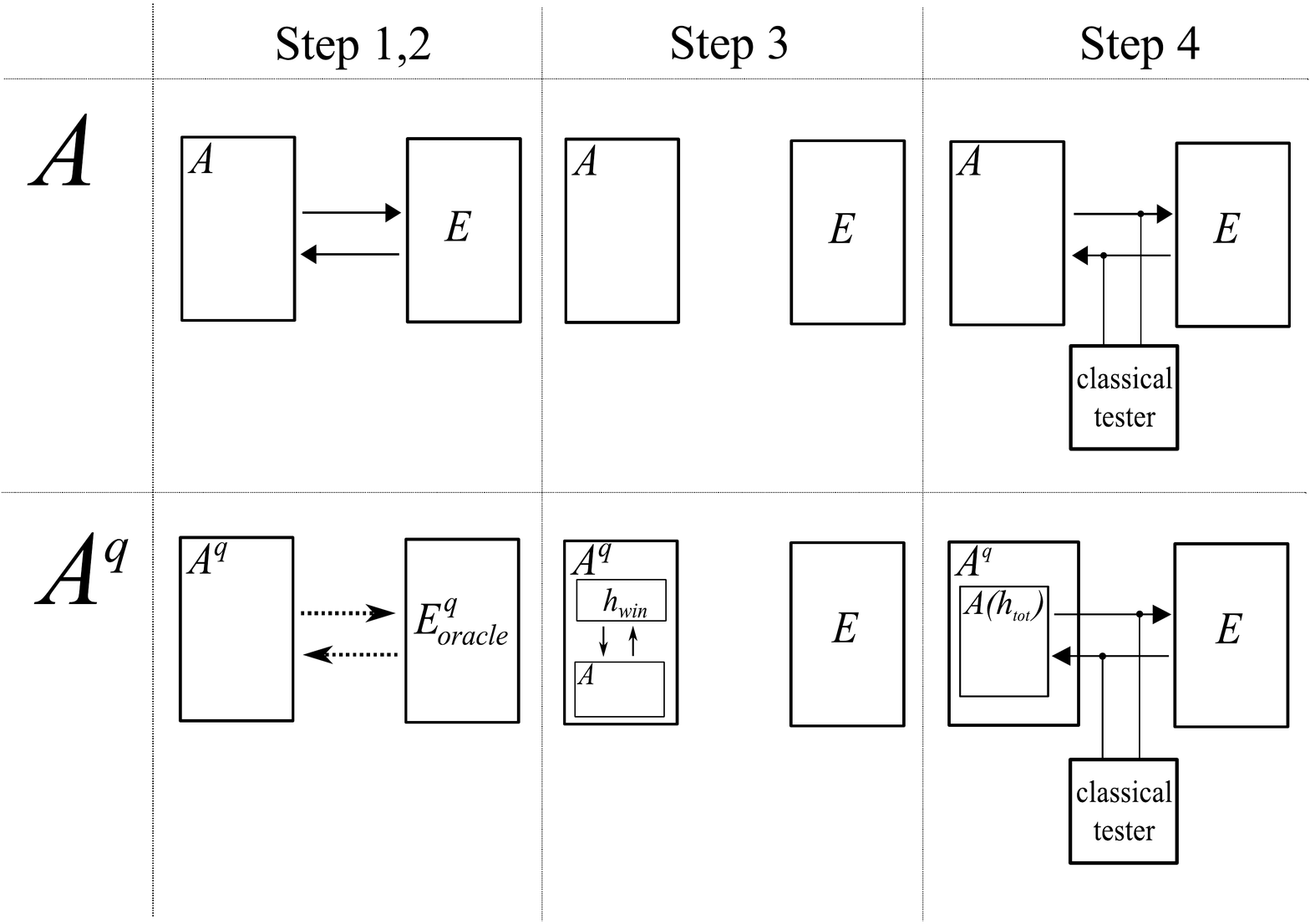}}

\caption{\label{construction}
The figure illustrates the differences between the agent-environment interaction for $A$ and the quantum-enhanced $A^q$. In Steps \ref{prep1} and \ref{prep2}, $A^q$ uses access to the oracular instantiation of $E,$ and obtains a winning sequence in, on average, a quadratically reduced number of interaction steps. At Step \ref{training}, $A^q$ simulates the agent $A$ internally, and `trains' this simulation to produce the sequence $h_{tot},$ derived from the winning sequence. During this time, there is no interaction between the agent and the environment, as this is `in between' time-steps. In Step 4, $A^q$ simulates $A(h_{tot})$ (using the obtained winning sequence $h_{win}$),  for the remainder of the interaction, now with the classical environment $E$. The interaction can be classically tested from this point on.
}
\end{figure}

This finishes our specification of the quantum-enhanced learning setting, and it is illustrated in Fig. \ref{construction}.

We now briefly clarify the purpose of the steps in the construction. The construction is designed to guarantee improvement in luck-favoring settings. Steps \ref{prep1} and \ref{prep2} simply utilize Grover-like search to obtain (at least) one winning sequence of steps in the given environment, in time quadratically faster than would be possible for a classical agent. 
To understand the rest of the construction, we can ignore the quantum aspects and consider how one could utilize the knowledge of an agent $A$ given a winning sequence, without specifying the internal model. Step \ref{training} aims to achieve just that - it simulates an interaction with the agent $A,$ and resets the agent, until the desired sequence has been achieved. In quantum information terminology, the runs of the agent $A$ get post-selected to the winning branch. However, the number of interactions that have been experienced to this point are ($k \times\sqrt{n^M}$ times) larger than the length of the winning sequence ($M$). To compensate for this, and to put $A$ and $A^q$ on equal footing, this `postselection' is iterated on a larger scale - until the agent (by chance alone) reproduces the winning sequence $k \times\sqrt{n^M}$ times in a row\footnote{Alternatively to this, one can consider a broader definition of luck-favoring settings, where the two histories $h$ and $\tilde{h}$ (experienced by the `lucky', and `unlucky' agent respectively) may be of unequal lengths.}.

This choice of the process of `training' a reinforcement learning model, given a winning sequence (or many winning sequences) is not crucial for our main point. However, regarding the optimization of the performance of the learning agent $A^q$, depending on how much is known about the learning model underlying $A,$ it should be chosen such that it maximizes the expected performance. We will further comment on this later.

To get further insight into the expected performance of $A$ versus $A^q$, consider the average configurations (relative to input-output behavior) of the agents $A$ and $A^q$ after the first $t$ steps.

Concerning agent $A^q$, after the time-step $t,$ and except with probability $O(\exp(-k))$, its behavior will be identical to the behavior of $A(h_{tot}),$ where $h$ is the history containing $(1+k\times\sqrt{n^M})$ successful move sequences glued together. 

%Due to the nature of Grover's search, the winning history $h$ is chosen uniformly at f from the set of sequences of winning histories of length $M_{max}$.

The configuration of the classical agent $A$, facing the same environment, is a bit more complicated, and what can be said is restricted by the fact that we do not specify the learning model of $A$.

Agent $A$ has also undergone $t$ interactions with the environment, that is, $(1+k\times\sqrt{n^M})$ complete games. 

The probability, however, of $A$ having seen at least one winning sequence (assuming there is no prior knowledge available to the agent) is upper bounded by the following expression:
\EQ{
P =1- \left( 1 - \dfrac{1}{n^M}   \right)\left( 1 - \dfrac{1}{n^M-1}   \right) \cdots \left( 1 - \dfrac{1}{n^M-k\times\sqrt{n^M}}   \right),
}
where we have taken into account the fact that the agent may (in the optimal case) never re-try a sequence which was not rewarded.
That expression further simplifies to 
\EQ{
P = \dfrac{k}{\sqrt{n^M}} + \dfrac{1}{n^M}
}
which decays exponentially to zero, for any fixed $k$, in $M$.
If we, for concreteness, set $k=M$ we have that both the probability $P$, and the failure probability of the quantum agent $O(\exp(-k)) = O(\exp(-M))$ decays exponentially  in $M$.
Thus, except exponentially small probability in $M$, the quantum agent will, from time-step $t$ onwards behave as 
$A(h_{tot}),$ where $h_{tot}$ has a maximal rate of rewards, whereas the classical agent will behave as   $A(h_{fail}),$ where $h_{fail}$ has not one rewarded percept.
Then, relative to any figure of merit $Rate$ which is increasing in the reward frequency (and depends only on the rewards) we have that 
$Rate(h_{tot}) > Rate(h_{fail})$.

Now, if the environment is luck-favoring, by Eq. (\ref{main-eq-luck}), from time-step $t$ onwards, the average performance of $A^q = A(h_{tot})$ will  beat the performance of $A(h_{tot})$ except with exponentially small probability, relative to the classical tester.

These observations form the first qualitative result:

\TH\label{th1}
Let $E$ be a controllable environment, over action space $\mathcal{A}$,  thus it is, on the agent's demand, accessible in the form $E^q_{control}$. Moreover, let $E$ correspond to a deterministic, fixed-time $M$, single-win game, with a unique winning sequence of length $M$, for the period of $O(|\mathcal{A}|^M)$ time-steps (after which it no longer needs to be controllable, nor deterministic, fixed-time, single win).
Let  $A$ be a learning agent such that $(E, A)$ are luck-favoring for all histories, relative to some figure of merit $Rate(\cdot)$, which is increasing in the number of rewards in the history, and which only depends on the rewards.
Then there exists a quantum learning agent $A^q$ based on $A$ which outperforms $A$ in terms of $Rate(\cdot)$ and relative to a chosen sporadic classical tester.  
\HT

The above is the least one can establish. If we start specifying the scenario further, by e.g. fixing the $Rate(\cdot)$ to be an effective (normalized) counter of the rewards, then we can also consider the average number of interaction steps which the classical agent needs to perform (relative to the quantum agents $t =   k \times\sqrt{n^M}\times{M}+M$)  before the two agents can even in principle start achieving approximately equal behaviors in terms of the rate\footnote{Note that every sensible learning agent will, given a sufficient number of steps, start producing the winning sequence every subsequent game. In this case, the rate will be maximal for all such agents.}.
As we have clarified, the classical agent requires an average $t_c \in  O(n^M\times{M})$ interaction steps (so $O(n^M)$ complete games), before a rewarded sequence is seen even once, on average. Thus this establishes a reasonable lower bound on the order of the number of steps required for a classical agent to start approaching the performance of the quantum agent. This constitutes a type of a quadratic improvement.
However, making such claims more formal requires further specifying the underlying learning model. In this paper we wish to establish more general claims, and leave more specific analyses for future work. 

{Nonetheless, for concreteness, we can list examples of learning models, and task environments, where the quadratic improvement mentioned above is easy to argue. 
In particular, we can backtrack to the basic maze example given in Fig. \ref{fig-mazes}. If we additionally label each percept (so each percept contains the arrow specifying short paths, and an identifier, which does not change the analysis thus far), then many well-studied reinforcement learning models (e.g. Q-Learning~\cite{2003_Russel}, Policy iteration~\cite{1990_Sutton} or the more recent Projective Simulation~\cite{2012_Briegel,2014_Melinkov} model), together with the maze environment (with a unique winning path) do form luck-favoring pairs for all histories, so Theorem \ref{th1} applies. Thus a speed-up in learning is possible. 
To further explain why this is the case (but without going into the details of these learning models), recall that in this single-win, bounded maximal time $(M)$ case, there is only one $M$-length history which has a reward. Moreover, a rewarding percept can only appear after exactly $M$ interaction steps, as the game is reset after each $M$ steps. Next, note that the length any history can be written in the form $l\times M + q,$ with $q<M$, for some integers $l, q$.
In such a history the last $q$ percepts cannot be rewarding, so we can focus on histories of lengths $l \times M$. This can be interpreted as an $l$-fold concatenation of histories of lengths $l$. Each one of these $l$ sub-histories either has exactly one rewarding percept, or does not, and it does only if that sub-history is the unique winning sequence.
In the learning models we have mentioned, applied to such an environment, for every game where a winning sequence has been executed, the probability of executing the same winning sequence can only increase. This implies that for any two histories (independently of their length)  Eq. \ref{main-eq-luck2} holds. Moreover, if the environment does not change, it holds for all execution lengths (as specified in the Definition \ref{lfa}), hence Theorem \ref{th1} does apply.
}

Going beyond the learning models we have mentioned, it is arguable that any learning model which is \emph{not} luck favoring with such a maze environment is a deficient learning model, as this would imply that the performance of the model (or the agent) does not monotonically improve as the agent encounters new short(er) paths.

In contrast, environments which are not luck favoring with standard learning models are possible to concoct. 
Simplest examples include malicious environments that change the rules
 depending on the initial success of the agent. In this case, having a low efficiency in the exploration phase may be beneficial in the long run, but such scenarios are quite artificial.
In the next section, we will consider further generalizations of environments where a speed-up is possible, and in the process touch upon more reasonable (and more general) settings where being lucky may be not as advantageous, and comment how to deal with such settings.

\section{Reducing the assumptions for quantum-enhanced agents}
\label{extensions}
We will first eliminate some of the assumptions we have introduced, for technical convenience only, in the previous section. 

In the set-up of our main result, we have assumed that the underlying classical learning model has a non-zero probability of hitting the winning sequence of moves, starting from the initial state. This assumption was necessary for the proposed `training' phase in Step \ref{training} of the construction, as we have demanded that the agent produces the winning sequence, without any prior history (and the agent was re-set otherwise).

This assumption will be true for most learning models which have a stochastic component, but, in extreme cases, the agent may be deterministic and output the possible sequences in a given order.
However, even in this case, recall that we assume that the full specification of the underlying learning model is known. In particular, we can (for the purpose of the construction of $A^q$) reverse-engineer under which conditions (modifications of internal parameters, whatever they may be) the classical learning model outputs the desired sequence, and simply enforce this scenario\footnote{For instance, for the (initially) deterministic example, we may simply internally let the agent run its course up until it does produce the winning sequence.}. This reverse-engineering may be extremely time consuming in general, but in the strictly synchronous model the internal times do not matter. 

Next, we have assumed that the game has only one winning path, and, related to this, we have assumed that the fixed time per game is $M_{max} = M$ equal to the (twice the) length of the winning path.
Removing these assumptions, for the most part, only leads to a quantitative difference in our analysis. 
Since the game is still assumed to be single-win, this will only imply that a certain fraction $f$ of the total number of action sequences available is winning, which is not just the inverse of the total number.
To clarify this, for the case where the maximal allowed time is larger, then it is possible that there are more winning sequences (\emph{e.g.} any one where the agent diverges from the optimal path for one step, but then returns to the route is also a winning sequence ).
But, Grover's search will work equally well with multiple winning sequences, and again yield a quadratic improvement.

Nonetheless, there is a subtlety occurring with respect to the underlying classical learning model, and the likelihood that a learning model, with such environments with multiple winning sequences, constitutes a luck-favoring pair.
In particular, when multiple rewarding sequences are possible, and available to the agent, the agent has a larger set from which to draw conclusions about the underlying structure. 
As constructed, the quantum agent will have found a winning sequence, which is then iteratively `force-fed' to the underlying learning model many (specifically, $1+k\sqrt{f^{-1}}$ many) times. At the face of it, the simulation of $A$ within $A^q$ sees the correct sequence many times, but it is always the same sequence. In contrast, if $f$ is not prohibitively small, the classical agent may see a s winning sequence a substantially smaller number of times, but the sequences it does see may differ.
If the underlying classical learning model is clever, the latter option may lead to better performance in the long term, which may violate the strict luck-favoring assumption.
A related problem is that the simulation $A$ within $A^q$ may also suffer from, in the language of machine learning, so-called \emph{overtraining} (also known as overfitting) - becoming specialized to handle this one sequence, and inapt to adapt to changes.
If the underlying learning model can suffer from overtraining (which we know prior to the design of the matching $A^q$), it may be overall better to `force-feed' the simulation of $A^q$ with a smaller number of copies of the winning path, and this should be tailored for the underlying learning model.

If it is also beforehand known that there are multiple solutions, and the underlying model is such that it benefits greatly from having access to multiple solutions, one can run the `Grover search phase' any desired number of times, instead of just once.
Note that the Grover's algorithm will output a winning sequence from an underlying uniform distribution over winning sequences. A-priori, there is a probability of waisting a search to obtain the same sequence again. But this is easily prevented - once (a set of) winning sequence has been found, the agent can, additionally to the phase flip induced by the oracle $E^{q}_{oracle}$, also phase-flip all sequences which are already found - this will effectively realize a search for just the un-found sequences.
In the scenarios such as described above, it is clear that the underlying model matters too much to make broad statements which we target in this work. Nonetheless, the considerations above give confidence that speeding up exploration using quantum interactions is useful in most realistic settings.

For the remainder of this section we will consider more radical extensions, where the environment is no longer deterministic, and when the games are not fixed-time, or, related to this, not single-win. To deal with such environments, we will introduce more general oracular instantiations of classically specified environments.
We begin with stochastic environments.
\paragraph{Stochastic environments}
\label{StochEnv}
In the case of stochastic environments, the output percepts are sampled from a distribution over a Cartesian product of the required number of percept sets.
It will be convenient to introduce a bit of notation. We will first consider fixed-time games, fixed at $t$ steps, in which case there is a mapping from $t$ actions to a distribution over $t$ percepts (we ignore the initial percept of the environment for simplicity), realized by the environment. This mapping $E_{in-out}$ we can write as:
\EQ{
\dm{a_1, \ldots, a_t} \stackrel{E_{in-out}}{\longrightarrow} \sum_{s_2, \ldots, s_{t+1}} P(s_2, \ldots, s_{t+1} \vert a_1, \ldots, a_t) \dm{s_2, \ldots, s_{t+1}} \nonumber,
}
where we have already introduced the bra-ket notation, for convenience, and have encoded the relevant probability distribution $P(s_2, \ldots, s_{t+1} \vert a_1, \ldots, a_t) $ into a mixed state.
The mixed quantum state, on the right-hand side of the expression above, can be purified by using a (at most) double-sized register:
\EQ{
\ket{\psi_{out}} = \sum_{s_2, \ldots, s_{t+1}} \sqrt{P(s_2, \ldots, s_{t+1} \vert a_1, \ldots, a_t)} \ket{s_2, \ldots, s_{t+1}}\ket{s_2, \ldots, s_{t+1}},
}
and in general $\ket{\psi_{out}}$ is highly entangled.
In the case the reward is issued (or withheld) at time step $t+1,$ the last percept $s_{t+1}$ encodes it, and some of the sequences of percepts end with a winning percept and some do not, for a fixed sequence of actions.

In this case, we cannot simply ignore the responses of the environment, if we wish to find the best possible action sequence. 
Nonetheless, there are still many ways to construct a meaningful oracle (or oracles) which will help find the best sequence of actions. Here we present one possibility.
For notational convenience, in what follows we will represent sequences of percepts and actions in bold, thus with $\textbf{s}$ and $\textbf{a}$, respectively.

First, we define the ``raw percept'' unitary oracle $U_{\mathcal{S}'}$, which, given an action sequence, produces the corresponding sequence of percepts, given as purifications. \EQ{
\ket{\textbf{a}} \otimes \ket{\epsilon \cdots \epsilon} \stackrel{U_{\mathcal{S}'}}{\longrightarrow} \ket{\textbf{a}} \sum_{\textbf{s}} \sqrt{P(\textbf{s} | \textbf{a})} \ket{\textbf{s}}\ket{\textbf{s}}. \label{Usubs}
} 
 In the above, and what follows, all percepts are raw, meaning in the unrewarded subspace.
The next oracle is a rewarding oracle $U_{R}$:
\EQ{
\ket{\textbf{a}} \otimes \ket{\textbf{s}}\ket{\textbf{s}} \stackrel{U_{R}}{\longrightarrow} (-1)^{R(\textbf{a}, \textbf{s})}\ket{\textbf{a}} \otimes \ket{\textbf{s}}\ket{\textbf{s}}, \label{Urew}
}
where $R$ is the binary reward function, depending on the actions and the realized percepts.

Access to these two oracles, and their inverses allow for a quantum amplitude amplification protocol (QAA) \cite{2000_Brassard}.
Recall that a QAA protocol allows for the preparation of a target state $\ket{\psi_{tar}},$ from some initial state $\ket{\psi_{init}},$
assuming the access to three types of oracles. First, a preparation oracle, generating $\ket{\psi_{init}},$ from some fiducial state. Next, we require two (unitary) reflection operators, $U_{init} = \mathbbmss{1} - 2 \dm{\psi_{init}}$ and $U_{tar}= \mathbbmss{1} - 2 \dm{\psi_{tar}}$ reflecting over the initial, and the target state, respectively.
Then, by the results of \cite{2000_Brassard}, the sequential application of $U_{init} U_{tar}$ to the initial state $\ket{\psi_{init}},$ on the order of $O( \vert\bra{\psi_{init}} \psi_{tar}\rangle\vert^{-1} )$ times, will yield a state within a constant distance from the target $\ket{\psi_{tar}},$ in terms of the fidelity.
For our case, the initial state is:
\EQ{
\ket{\psi_{init}} = \sum_{\textbf{a}, \textbf{s}} \sqrt{P(\textbf{s} | \textbf{a})} \ket{\textbf{a}}    \ket{\textbf{s}}\ket{\textbf{s}},
}
appropriately normalized, which just depends on the length of sequences and sizes of the action space.
It will be convenient to represent the conditional distribution above as a joint distribution, so $P'(\textbf{s} ,\textbf{a})=P(\textbf{s} | \textbf{a}).$
The target state is then given with 
\EQ{
\ket{\psi_{tar}} = \sum_{\textbf{a}, \textbf{s}} \sqrt{P'(\textbf{s} , \textbf{a}\vert R(\textbf{s},\textbf{a})=1)} \ket{\textbf{a}}  \ket{\textbf{s}}\ket{\textbf{s}},
}
again up to renormalization. Note that we have conditioned the joint distribution $P'(\textbf{s} ,\textbf{a})$ on those instances where the action-percept sequence yields a reward.
The measurement of the action component of this state will yield an action sequence given by the following probability:
\EQ{
P(\textbf{a}) = \sum_{\textbf{s}} P'(\textbf{s} , \textbf{a}\vert R(\textbf{s},\textbf{a})=1),
}
which is the just \emph{expected} reward of the sequence of actions $\textbf{a}$. This sampling favors the best sequence of actions, that is, those which have the highest expected reward, and completely suppresses actions which cannot yield a reward.

What remains to be show is how the reflectors  $U_{init}, U_{tar}$ are realized by the oracles we have, and how the initial state is prepared. The preparation of the initial state is obtained by running the oracle $U_{\mathcal{S}'}$ on a uniform superposition of action states, see Eq. (\ref{Usubs}). 
While we cannot construct a true reflection over this initial state, we can construct a reflection which performs the same on the subspace we will be working with (the \emph{busy subspace}).
Define $U_{init}'$ as applying $(U_{\mathcal{S}'})^\dagger,$ followed by the reflection over the uniform superposition of the action sequences, applied to the action register. Symbolically, given the normalized state ${\displaystyle \ket{\phi} = 1/\mathcal{N} \sum_{\textbf{a}} \ket{\textbf{a}}}$ we have
\EQ{
U_{init}' = \Big( \underbrace{\left( \mathbbmss{1} - 2 \dm{\phi}\right)}_{\textup{Action\ register}} \otimes \mathbbmss{1} \Big)      (U_{\mathcal{S}'})^\dagger.
}
Note that $U_{init}'$ behaves like $U_{init}$ on the subspace $\textup{span} \{   U_{\mathcal{S}'} \ket{\textbf{a}}\ket{\epsilon, \ldots,\epsilon}   \}_{\textbf{a}}$.

The reflection $U_{tar}$ is similarly realized, on the same subspace, by the reward oracle $U_R$.

\paragraph{Environments with multiple reward values and without fixed times}
In the case of a multiple reward setting, or similarly, in the case where rewards are not just binary, the simplest solution is to design oracular instantiations of environments, where the sum of total rewards per history is represented.
In this case one can, out of such oracular instantiations, construct other oracles where the reward is specified by a threshold function applied to the actual cumulative reward. Then, we can quantum search for the highest reward if the exact highest reward is known. If only a bound is known, then we can resort to a binary search procedure, which will yield only a logarithmic overhead.
However, in the case of multiple rewards, being greedy, and targeting only for the highest reward possible may be suboptimal. One can consider settings where histories of $r-1$ reward value are plentiful (hence quick to find) whereas maximal rewards $r$ are scarce, or unique. Then it may make sense to optimize the reward value we are looking for, and this then becomes learning model, and environment specific. We leave this optimization problem for future work.

In the setting of epochal (or game) environments, it is arguably equally natural to consider settings where the game ends with the first win, as it is to consider fixed times.
Even in the case of first-win games, it is possible to specified oracles which are defined for a sequence length $M_{limit}$ with upper bounds any ``reasonable'' interaction length. Such oracles would, for instance, take any sequence length, and append the fiducial state until length $M_{limit}$ is obtained. 

The final extension we consider here removes the epochal structure from the environment - this constitutes the most general environments. Recall, epochal structure was equivalent to stating that only a suffix of the elapsed history (that is, only a certain number of the most recent interaction steps), rather than the entire history, specifies the maps of the environment. The cases we considered were where $t-$step suffixes matter - which correspond to fixed-time games - and suffixes up to the last reward, which corresponds to first-win games.
There are easy generalizations of the two scenarios above which reduce to the settings we have considered, for instance, where obtaining a particular percept from the environment (but irrespective of the reward status) signifies a re-set.  
More involved generalizations include  environments, where there is a well-defined period at which the environment is re-set, but is also slightly altered. This would correspond to `slowly evolving oracles', in which case, whether or not similar techniques to the ones presented could be used, would greatly depend on the rate of change.
Finally, in the most general case, there is no straightforward mapping from environments to useful oracles. However in such environments, which effectively arbitrarily change at an arbitrary rate, it becomes questionable whether any learning model could learn anything, as it is not clear whether there is anything (any fixed parameter) to be learned at all.
Such, and considerations of other possible extensions are beyond the scope of this work.

\section{Building indirect oracular access to environments}
\label{indirect}

In the previous sections we have considered quantum extensions of classical environments, and have shown that, in general, the quantum extensions may disallow any advantageous quantum interaction. 
Then, we have extended the scope of the quantum environments we consider.  We have included particular oracular instantiations of the given classical specifications, which were tailored for a type of a quantum enhancement. In particular, we sought to expedite the typically necessary exploration phase of learning, which precedes the exploitation phase.

In this section, we will, through characteristic examples, re-examine the true quantum extensions of an otherwise classically specified environment, and examine what kinds of oracles they natively make. For clarity, we will begin by considering particular classes of classical task environments which will then highlight  specific problem points, which we have already sketched. 
Recall that in all but the simplest cases (considered later) the key problem lies in the fact that the environment must keep (copies of) the actions in its internal memory, but also in that the perceptual responses themselves prohibit the direct use of the environment as an immediately useful quantum oracle.

To circumvent this problem, and enable the agent to achieve an effective oracular access, we will relax the model of interaction, and grant additional powers to the agent.
In particular, we will assume that, at least at particular time-step intervals, the agent has access to the register (memory) of the environment, or to the purifying systems of the environment (if these exist). We will call these two options \emph{register hijacking}, and \emph{register scavenging}, respectively.

%\begin{figure}
%\caption{Register scavenging and hijacking}
%\end{figure}

Before this, we briefly reflect on what level of violation of the integrity of the environment hijacking and scavenging actually constitutes, on a conceptual level. We begin with these consideration as the approach we present is, ultimately, specific to the particular model for environments we have proposed in this work.

In the model we have presented, an environment comprises, in general, four conceptually different parts: specification, memory, interface and its own local environment (purifying registers).
With \emph{specification} we refer to the description of the physical system which determines and realizes CPTP the maps which define the agent. \emph{Memory} pertains to the internal register of the environment over which the environment has complete control - this memory can, for instance, record the histories of interaction with a particular agent. The \emph{interface} is the component of the environment which an agent has access to and can arbitrarily manipulate. Finally, the \emph{purifying registers} are systems which the environment discards or ceases to control in between its  activities\footnote{Whether or not purifying systems necessarily exist in a physical sense, strikes at the heart of some of the hardest problems in foundations of quantum mechanics.
It is true that for any environment, specified by CPTP maps, we can construct an indistinguishable environment which acts unitarily on its (enlarged) register. Whether or not such purifying registers, and dilated unitary evolution necessarily physically exists (and for scavenging to make sense, they must, as otherwise the agent would have nothing to scavenge anyway) is easily seen to be a generalization of the age-old question whether the Universe itself is in a pure state, and if it evolves unitarily. We do not aim to address these questions to any extent in this work, and our statements about purifying systems being constituent elements of an agent/environment (albeit not parts under active control) should be taken as an assumption we are making, rather than a statement about quantum mechanics. If the universe is such that CPTP maps can exist without being parts of an actual underlying unitary evolution (such as \emph{measurements} are sometimes taken to be) then environments defined by such maps clearly cannot be scavenged, even in principle. 
This issue reflects the flip-side of the problem which occurs in general cryptographic settings -- one must typically assume that for every (mixed) state a legitimate party has, the adversary has access to its purification.
}. 

The specification of the environment, above all other components, must be outside the agent's influence, as it is what a-priori defines the environment, and what it rewards. The other components actually pertain to \emph{how} the environment does what it does.
 The scavenging of discarded registers seems to be the smallest infraction, as the state of these does not influence the behavior of the environment in the future\footnote{This  statement deserves a caveat. It is well known that unitary evolutions over two systems, where the joint input state is correlated, do not necessarily yield a CPTP evolution on the subsystems. In this sense, inputting scavenged states back to the environment may influence the effective maps realized by the environment. The using of this fact, however, may be considered to be in the legitimate toolbox of the scavenging agent, as the defining physical maps of the environment are unperturbed. }. From a practical perspective, scavenging may constitute a serious challenge for the agent, say in the case of macroscopic environments, or it may be comparatively straightforward, if the environment is a part of a quantum system in a quantum lab. Register hijacking corresponds to the capacity of the agent to alter the memories of the environment. While being manifestly invasive, it still may be acheivable in computational, or laboratory settings.
 
Note that both environments, and agents, could have also been defined differently than what we have presented in this work - for instance by using a unique register, which also contains the specification of all maps which, in our paradigm define the agent and one, universal map (driven by one specified Hamiltonian), applied, sequentially on the system register. In this context, altering the `program' part of the memory would constitute the altering of the agent's characteristic maps. Such an alteration is what we have effectively done when we simply assumed we had access to an oracular instantiation $E^q_{oracle}$ of a given environment $E$. Here, we wish to disallow this, and consider what can be done if we assume less - access to just the `work register' of the environment, which, arguably, does not constitute a genuine change of the environment.

Now we begin with our first example.\\

\noindent\textbf{Example 1} Consider a simple environment $E$ in which the raw percept space $\mathcal{S}'$ is trivial (contains one element ``$\ast$"), and the reward space $\Lambda = \{ 0,1 \}$ is binary. Thus, the complete percept space $\mathcal{S}=\mathcal{S}' \times \Lambda$ is given by $\mathcal{S}= \{ (\ast, 0), (\ast, 1) \}$. The action space is given by the set $\mathcal{A} = \{a_i \}$, and the environment rewards some of the individual actions, specified by a (non-trivial) reward function $R: \mathcal{A} \rightarrow \Lambda$.
Thus, given an action $a_i$, the environment responds with the percept $(\ast, R(a_i))$, and the environment requires no memory.
This is perhaps the simplest non-trivial environment, and matches the definition of a standard boolean-function-specified oracle.

Now, we consider the possible extensions of $E$ to the quantum domain, that is, the elements of the set $\mathcal{E}_c(E)$. 
We will separately analyze the communication, and the embodied model as, specifically in this scenario, they offer different possibilities.
\paragraph{Communication model}
Recall that the quantum environment $E^{q} \in \mathcal{E}_C(E)$ must, by definition, match the outputs of $E$ in the presence of a classical tester.
This presents certain restrictions on the possible quantum maps of the environment. Let $\mathcal{F}: \mathcal{L}(\mathcal{H}_C) \rightarrow  \mathcal{L}(\mathcal{H}_C) $ be the map effectively realized on the register $R_C$ by the action of the environment $E^q$. The constraint that $E^{q} \in \mathcal{E}_C(E)$ implies that this map must satisfy:
\EQ{
\mathcal{F}(\dm{a_i}_{R_C}) = \dm{\ast, R(a_i)}_{R_C}. \label{env-map-1}
}
This will already imply, for instance, that whenever $|\mathcal{A}|>2$, the map $\mathcal{F}$ cannot be reversible and, in particular, some information about the input is lost (transferred to the environment). It can also be shown that, for general reward functions $R,$ and settings with many percepts, such a map cannot maintain superpositions, without perturbing relative weights, as this would compromise linearity.
While this does not imply that a map satisfying Eq. (\ref{env-map-1}) (and the environment realizing it) does not allow the generation of genuinely quantum states by the environment (and hence a non-trivial quantum interaction), it certainly does not match the definitions of standard quantum oracles we have given before, and which we target here.
\paragraph{Embodied model}
In the embodied model, the situation is slightly improved as  the existence of two interface registers effectively allows the agent (environment) to have access to one state from the previous (half of a) time-step - its output action.
We remind the reader that there are two classical testers one may consider in the embodied setting. The strong classical tester, which models complete dephasing at the interfaces, and the standard tester, which copies only the interface of the system (agent or environment) which acted last. In the case of the strong tester, we essentially get the same situation we had in the communication model. Thus, we shall consider the standard tester. 
The map realized by the environment on the registers $R_{I(A)}R_{I(E)}$, at some time-step, in general given with
 $\mathcal{F}:~\mathcal{L}(\mathcal{H}_{\mathcal{A}} \otimes \mathcal{H}_{\mathcal{S}}) \rightarrow  \mathcal{L}(\mathcal{H}_{\mathcal{A}} \otimes \mathcal{H}_{\mathcal{S}}).$ 
So, at least the reversibility (and unitarity) is not immediately prevented by the dimension mismatch between the domain and codomain.
This map must satisfy:
 
\EQ{
\mathcal{F}\left(\dm{a_{(t)}}_{R_{I(A)}}  \otimes \dm{\ast, R(a_{(t-1)})}_{R_{I(E)}} \right) = \nonumber\\
\rho_{R_{I(A)}} \otimes \dm{\ast, R(a_{(t)})}_{R_{I(E)}}. \label{env-map-2}
}
In the above expression, the subscripts clarify from which time-step the action comes -- note that the environmental interface, before the subsequent move of the environment, may hold the reward for the previous action $a_{(t-1)}$. This depends on the maps of the agent, who could have influenced the state of that register.
The right hand side of the equation (after the map of the environment) must be such that the environmental register contains $R(a_{(t)})$ -- the reward for the current action $a_{(t)}$, and we the allow any state for the register $R_{I(E)},$ after the action of the environment.
The map above does allow that the environment effectively implements a type of an oracle which is known to be useful. Specifically, if the environment always first resets its interface to one fixed state, say $\ket{\ast, 0},$ after which it acts unitarily, a map $\mathcal{F'}$ which satisfies Eq. (\ref{env-map-2}) is given as follows:
\EQ{
\mathcal{F'}\left(\ket{a_{(t)}}\bra{a'_{(t)}}  \otimes \rho \right) =\ket{a_{(t)}}\bra{a'_{(t)}}  \otimes \ket{\ast, R(a_{(t)})}\bra{\ast, R(a'_{(t)})} \label{env-map-3},
}
for any state $\rho$.
This matches the specification of the quantum oracle for the oracular function $R$ given in Eq. (\ref{QFTor}) which has been advantageously used in certain quantum algorithms, as discussed previously.
We believe this result may have potential utility. However it is insufficient for our purposes as we are interested in attaining the stronger oracles given in Eq. (\ref{x-or}) (bit-flip oracle) and Eq. (\ref{z-or}) (phase-flip oracle).

Note that the constraints on the map $\mathcal{F}$ we have stated above are inconsistent with the specification of the the bit-flip oracle - the state of the register $R_{I(E)}$, after the action $a_i$ of the agent must contain $\ket{\ast, R(a_{(t-1)})}$ regardless of the initial state of this register. In non-trivial cases of the environment, with a non-constant function $R$ this implies irreversibility.
It is perhaps even more obvious that the specification of the phase-flip oracle is inconsistent with the classical specification of the environment -- the phase-flip oracle only changes the global phase of classical states, which cannot be detected by a classical agent, or a classical tester.
However, for these simple cases, the desired oracles can be generated, through register hijacking and scavenging.

\paragraph{Solution for the embodied model}
The embodied model allows for the simplest solution. Note that we have stated that the environment may be implementing the map given in Eq. (\ref{env-map-3})
provided the environment, as its first step in the realization of the map, resets its interface to, say, the $\ket{\ast, 0}$ state.
More precisely, the map $\mathcal{F}$ can be realized by the composition of a CPTP contraction to the state $\ket{\ast, 0}$  applied on the register $R_{I(E)},$ followed by a controlled unitary map, which, conditional on the state of the first register being in the set (more precisely, the corresponding subspace) $R^{-1}(1),$ rotates 
$\ket{\ast, 0}$ to $\ket{\ast,1}$, and does nothing otherwise.
In particular, we can choose this (controlled) local rotation to have eigenvalues $1$ and $-1$ (effectively, a Pauli-X gate on the relevant subspace), while maintaining the correct classical limit.
This assumption will later allow us to generate a `phase-kick back,' which transforms bit-flip oracles to phase-flip oracles -- to see this, note that if $U\ket{\ast, 0} =\ket{\ast,1},$ where $U$ is acting on a two dimensional space, and $U$ has a (-1) eigenvalue, then $U(\ket{\ast, 0} - \ket{\ast,1}) = - (\ket{\ast, 0} - \ket{\ast,1}) $.

However, we have stipulated that the environment contracts the state of its register to $\ket{\ast, 0}$. Here, we utilize a form of register hijacking.  One way to unitarily realize a contraction to the state $\ket{\ast, 0}$, for the register $R_{I(A)}$ is to prepare $\ket{\ast, 0}$ in the  register $R_{prep},$ which is a part of the entire memory register $R_E$ of the environment, swap the states of $R_{prep}$ and $R_{I(A)}$, and then if needed discard (trace out) $R_{prep}$. If the agent has access to $R_{prep},$ after the initialization of the environment to $\ket{\ast, 0}$, then it can substitute $\ket{\ast, 0}$ with $\ket{\ast,-}=\dfrac{1}{\sqrt{2}}(\ket{\ast, 0} - \ket{\ast,1}) $. This then turns the map of the environment into the phase oracle as desired. 
The same results can be obtained by assuming the agent can alter the state of $R_{I(A)}$ after the environment has re-initialized it, but before the controlled map has been applied. 

In the case of the communication model, similar can be achieved by further utilizing register scavenging.
Note that we can assume (while maintaining the correct classical specification of the environment) that the map realized by the environment, in the communication model is exactly the same as in the embodied model, which we have presented above, where the role of the \emph{agent's} interface is taken by an internal register of the \emph{environment}. To see this, just assume that the environment, as its first map, moves the contents of the $R_C$ register to an internal register. Then, this internal register can play the role of $R_{I(A)},$ and we can assume that it is also discarded after the map $\mathcal{F}$ has been applied to it and $R_C$.
Through register scavenging, the agent can collect back the discarded register. This then reduces the communication setting to the embodied setting.

As we have demonstrated with the example above, our strategy is to employ register hijacking and scavenging, to allow the agent to modify the type of effective maps the environment induces on the actions the agent outputs. In particular, we aim at effectively constructing the oracular instantiation  $E_{oracle}$ of the environment, as specified specified in Fig. \ref{fig4} (a), and explained in the corresponding section. Once this is achieved, the learning enhancements follow by the results of section \ref{lfa} in for luck-favoring settings.

Recall that we have shown earlier the particular quantum extension of the otherwise classically specified environment may a-priori prohibit any kind of quantum enhancements, and this remains true even if the agent has limited capacity to alter the internal register of the environment -- whatever the agent does will fail to yield any enhancements if, for instance, the environment always irreversibly de-phases the communication interface(s).
Thus, we will be concerned with the question of what legitimate quantum extensions can we choose which do allow enhancements, through hijacking and scavenging. 
We will briefly address the question to what extent our assumptions on how the environment is realized (what quantum extension the agent has access to) can be relaxed later.

\vspace{0.5cm}

\begin{figure}[h!]

Agent-environment $M-$step interaction:
\\
\vspace{0.5cm}
\EQ{
\Qcircuit @C=1.3em @R=0.8em {
&&\lstick{\ket{a_1}}&\qw& \multigate{5}{\mathcal{U_R}} & \qw &\qw&\rstick{\ket{s_2}} &\\
 & &  \lstick{ \raisebox{0.6em}{ \vdots} \hspace{0.7em}   }         &{/}  \qw &\ghost{\mathcal{U_R}}& {/} \qw&\qw  & \rstick{\hspace{0.3em}\raisebox{0.6em}{ \vdots}    } &  \\
&&\lstick{\ket{a_M}}      &\qw& \ghost{\mathcal{U_R}} & \qw &\qw&\rstick{\ket{{s_{M+1}, \lambda}}} &\\
&&\lstick{\ket{\epsilon}} &  \qw  & \ghost{\mathcal{U_R}}  & \qw &\qw&\rstick{\ket{a_1}} & &  \\
& &\lstick{ \raisebox{0.6em}{ \vdots} \hspace{0.7em}   }         &{/}  \qw &\ghost{\mathcal{U_R}}& {/} \qw&\qw  & \rstick{\hspace{0.3em}\raisebox{0.6em}{ \vdots}    } & \\
 &&\lstick{\ket{\epsilon}} & \qw  & \ghost{\mathcal{U_R}}  & \qw &\qw&\rstick{\ket{a_M}}&&\push{} \gategroup{4}{1}{7}{10}{1.5em}{..}\\
 & &  &   & &\hspace{-1.3cm}\textup{Environment }& && &\push{} \gategroup{4}{1}{7}{10}{1.4em}{..} \gategroup{4}{1}{7}{10}{1.3em}{..}\\
 \\
 \\
 }\nonumber
}

\caption{\label{fig5}   Illustration of an n-step interaction, for deterministic, fixed-time, single win games. The $\lambda$ symbol, appended to the percept label $s_{M+1}$ designates that this particular percept state is rewarding ($\lambda=1$) or non-rewarding ($\lambda=0$).}
\end{figure}
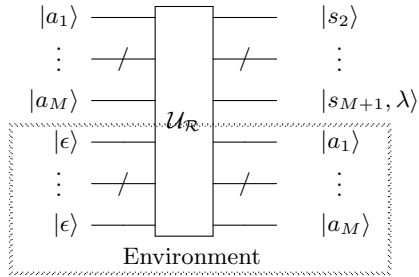

\noindent\textbf{Example 2.} Next example is more involved, and highlights the problem which occurs when the environment offers multiple percepts. 
In particular, we will consider deterministic fixed-time single-win games, introduced in Section \ref{good-oracles} and further elaborated in Sections \ref{main-section} and \ref{lfa}.
For this particular case, we have already shown what the map of the agent realizes, from the perspective of the agent, as represented by the circuit given in Fig.~\ref{fig5}.

In Fig. \ref{fig5}, the sequential inputs, actions $a_k$, of the agent are depicted to be given in parallel. The output given by the environment to the agent (percepts $s_k$) are also represented to be output in parallel.
A brief comment is in order. Since the inputs of the agent are output unchanged by the environment, on the right hand side of the circuit, this implies that the maps of the environment are of a controlled form, relative to the classical basis, controlled by the agent's choice of the output actions. This is a restriction, as in general, the environment can apply any types of maps, which also modify the action sequence in its register. Nonetheless, any classically specified environment has a quantum extension where the maps of the environment have this form, and we shall consider this case. Fixing this aspect of the specification of the quantum environment, however, does not yet solve all the problems, and we will constrain the allowed specifications further. 
Recall that our goal is to find quantum specifications which allow us to, through register hijacking and scavenging, achieve the same input-output mapping we gave in the definition of the oracular instantiations $E_{oracle},$ specified in Fig. \ref{fig4} (a), and explained in the corresponding section.  

The steps of the construction are as follows. 

First, we show that we can find legitimate quantum extensions for which we can effectively realize the \emph{phase-kick back map}
\EQ{
\ket{a_1, \ldots, a_M} \ket{\epsilon, \ldots,\epsilon} \ket{\epsilon,\phi^-} \rightarrow (-1)^R \ket{a_1, \ldots, a_M} \ket{s_2, \ldots, s_{M}} \ket{s_{M+1}, \phi^-},
\\}
where the states $ \ket{\epsilon,\phi^-}$ and $\ket{s_{M+1}, \phi^-}$ will be specified later. Note that the map above is not a useful oracle per se, due to the percept states which appear on the right hand side, and which may be correlated to the actions.
Following this, using a similar construction, we can achieve the \emph{raw percept map}
\EQ{
\ket{a_1, \ldots, a_M} \ket{\epsilon, \ldots,\epsilon} \ket{\epsilon, \phi^+} \rightarrow  \ket{a_1, \ldots, a_M} \ket{s_2, \ldots, s_{M}} \ket{s_{M+1},\phi^+}. \\
}

As the final ingredient, we show how the environment, that is a particular legitimate quantum extension of an otherwise classically specified environment, can be chosen such that the maps of the environment are self-reversible.

 Assuming that the agent can manipulate the environment such that these maps are implemented, the procedure is as follows.
 
1) The agent first realizes the phase kick-back map. 

2) Next, it implants the percept-containing systems into the register of the environment, at the appropriate position -- this constitutes another instance of register hijacking. 

3) Finally, the agent engages in an an interaction which realizes the raw percept map.

Now, since the maps of the environment are chosen such that they are self-reversible, what is realized in total is easily seen to be equivalent to the desired oracular instantiation of $E_{oracle}.$

\paragraph*{Realizing the phase kick-back map}
The idea behind the specification of the environment which allows the realization of the phase kick-back map is essentially the same as in the simple case we had considered in Example 1. However, since, in this more general setting, we have a multitude of possible last-step percepts to consider. To handle this, we will assume that the percept space structure is of a tensor product form $\mathcal{H}_{\mathcal{S}}=\mathcal{H}_{\mathcal{S}'} \otimes \mathcal{H}_{\Lambda},$ where the reward is binary. Thus percepts of the form $\ket{s}\otimes \ket{0}$ span the non-rewarding subspace and the percepts of the form $\ket{s}\otimes \ket{1}$ span the rewarding subspace. Next we assume that, by default, all the percept states the environment generates are non-rewarding, and in order to make a given percept rewarding, the environment applies, additionally, a unitary map, mapping $\ket{0}$ to $\ket{1},$ which acts only on the (2-dimensional) reward-status-specifying Hilbert space. In particular, we can then choose this map to also be Hermitian, hence self-reversible. 

In this case, regardless of the actual percept the environment issues as the final/reward carrying percept, by hijacking of just the reward component, the agent can implant the state $\ket{\phi^-}= 1/\sqrt{2}( \ket{0} - \ket{1})$ in the place of the default $\ket{0}$ state, which achieves the kick-back. 
Moreover, all the maps the environment applies can be chosen to be of a particular controlled form. More precisely, we can assume that at each time-step $t$ the environment applies the map $U_t$ specified with
\EQ{
U_t \ket{a_1, \ldots, a_t} \otimes \ket{\epsilon}  = \ket{a_1, \ldots, a_t} \otimes U^{a_1, \ldots, a_t} \ket{\epsilon}.
}
The controlled maps $U^{a_1, \ldots, a_t}$ need only rotate $\ket{\epsilon}$ to $\ket{s_{t+1} (a_1, \ldots, a_t)},$ where we have made it explicit that the percept $s_{t+1}$ depends on the previous actions. The rewarding maps (flipping the reward bit from 0 to 1) can be assumed to be of the same form\footnote{In a deterministic environment, we can assume that the reward function value depends only on the actions of the agent. If the environment is stochastic, however, the environment must, in general, keep copies of the output percepts, as the reward function may depend on them as well.}.

Although each $U^{a_1, \ldots, a_t}$ acts on an $|\mathcal{S}|$ - dimensional Hilbert space, we only need to specify its action on the fiducial state $\ket{\epsilon}$, and we are free to choose the rest of the specification. But then, we can also choose it such that each $U^{a_1, \ldots, a_t}$ is Hermitian (and acting non-trivially only on the relevant two-dimensional subspace), hence self-inverse.
\paragraph*{Realizing the raw percept map} 
To achieve the raw-percept type of mapping, we follow exactly the same reasoning as for the phase kick-back map, with the difference that the agent implants the state $ \ket{\phi^+} = 1/\sqrt{2}( \ket{0} +\ket{1})$ in the reward carrying subsystem, instead of $\ket{\phi^-}$, as this state is invariant under the reward map. Note that the latter holds since all the maps the environment implements are Hermitian, which we have required for the self-reversibility. 
\paragraph*{Putting it all together}
To realize $U_{oracle},$ the agent first implements the phase kick-back map with the environment, for the price of one complete game. Thus, for a given choice of actions $\ket{a_1, \ldots, {a_M}},$ the agent obtains the state  $(-1)^{R} \ket{a_1, \ldots, {a_M}} \ket{s_1, \ldots,s_{M}} \ket{s_{M+1 }, \phi^{-}}.$
The percepts $s_k$ may depend on the choice of actions, but the state $\ket{\phi^{-}}$ does not, hence, even if the actions are given in superposition, this subsystem is not entangled to the rest of the system. 

Following this, the agent hijacks the register of the environment, and implants the subsystem containing the state $\ket{s_1, \ldots,s_{M}} \ket{s_{M+1 }, \phi^{+}}$ into the register of the environment, where the agent has substituted $\ket{\phi^-}$ with $\ket{\phi^+}$. Note that at this point, the agent's register (containing the actions part) and the environment's register may be highly entangled. Now, the agent completes a second complete game with the environment. Since the maps of the environment are self-inverse, and since they are non-trivially acting only on distinct registers, the maps of the environment implement an inverse of the raw percept map. This decouples the actions-carrying subsystems, from the percept-carrying subsystems.
The net result is the agent ending up with the state $(-1)^{R} \ket{a_1, \ldots, {a_M}} \ket{\epsilon, \ldots, \epsilon} \ket{\epsilon, \phi^{+}}.$
Since the register containing the fiducial state is not dependent on the actions, the agent has successfully realized the $E_{oracle}$ map as desired.

The total cost of interactions with the environment we used was $2 M$, as two full games were played. However, a fair counting of interactions should also count the communication cost of scavenging and hijacking.
Taking the modification of one percept/action containing register as to cost as much as one interaction step, we obtain $5M+1$ as a total count, so still on the same order, guaranteeing efficiency. 

%In the approaches we have given above, we have assumed that the quantum extensions of the classically specified environment we have given access to are, for our purposes, the most useful ones. \red{this part is now a repetition of the clarification from the beginning of the example.}
%However, certain assumptions on the extensions had to be made either way, as we have shown, there exist legitimate extensions which completely prohibit any quantum enhancement.
It is worthwhile to note that many other types of quantum extensions would be equally useful (if the type of maps is known). As an example, the simplest extension includes all maps which differ from the ones we chose by a local (known) unitary. But this observation can be further extended. In particular, if the environment is fully unitary, and the quantum specification of the controlled maps is known, the agent can always, in principle, recover the oracular map through hijacking and scavenging. 
%To avoid possible confusion, note that, naturally, we do not assume that the information on the reward map is known to the agent. \red{finish off this somehow}

We end this section by noting that hijacking and scavenging can be equally useful in more general environmental settings. In particular, approaches used in section \ref{StochEnv}, can be used in conjunction with hijacking and scavenging to significantly extend the class of environments which can be indirectly oracularized, provided suitable quantum extensions are available.

\section{Application - Model-based agency}
\label{mod-based}
In the previous section we have circumvented some of the problems which arise when one considers utilizing quantum access to quantum extensions of classically specified environments. To achieve this we have assumed that one has access to oracularized instantiations, which capture the relevant properties of the environment.
This assumption can be immediately fulfilled if the environment in question is actually constructed by the agent itself, and this is a part of the learning process in \emph{model-based} learning \cite{SuttonBarto98}. 
Roughly speaking, in model-based approaches to learning, the agent internally constructs a representation of the environment based on what it has experienced thus far.
This representation may, to a lesser or greater extent, be an explicit simulation of the environment, but in all cases the basic idea is the same: by interacting with the simulation the agent can establish optimal courses of actions, and plan, without incurring communication costs with the real environment. 
The validity of the conclusions the agent draws, inevitably suffers from a trivial bottleneck - the projection the agent makes is only as good as the simulation it has. However, the simulation is only built up through interactions with the environment.

Nonetheless, many of the successful and powerful learning models are within the paradigm of model-based learning \cite{2003_Russel}, and use the environmental simulations with success.
It is worth mentioning that, even the arguably most powerful methods for learning, in the sense of what types of environments they are designed to tackle, e.g. so-called Universal Artificial Intelligence approaches \cite{Hutter:07aixigentle} (which are not only computationally exorbitantly costly, but uncomputable in the strongest variant) do fit in the paradigm of model-based learning. Other approaches, including the Projective Simulation (PS)~\cite{2012_Briegel} approach, developed by one of the authors and collaborators, while, technically not being an explicit model-based formalism, is fundamentally based on notions of episodic memories, which capture the elapsed interactions with the environment. This memory is then queried in the search for the best actions. Such memory could be argued to construe the environmental model, in a broad sense.
One advantage of the PS model is that it has been successfully quantized, by some of the authors and collaborators, and quadratic improvements in deliberation times have been obtained~\cite{2014_Paparo}.  Improvement in deliberation times was argued to lead to a more successful agent in \emph{active learning}, which corresponds to the weakly synchronous model we explain presently.
Before this, we highlight that the results of this paper imply that \emph{all model-based learning models, which use an explicit representation of the environment, can be beneficially quantized}.
All that is required for this is to, instead of internally constructing a classically specified replica of the environment, one constructs the right oracularized instantiation.
It is correct to note that in the cases where the environmental simulation boils down to a simple lookup table, the same result could be achieved by directly ``Groverizing'' the search for the best actions. However, the environmental simulations may be a lot more elaborate \cite{Hutter:07aixigentle}, in which case it may be advantageous to have a near-generic approach for the beneficial quantization of the internal environmental simulation, as we have described in this paper.

As we have clarified earlier, in the setting where the interaction between the agent and environment is classical (for whatever reason), no quantum improvement is possible in the defined synchronous mode, where only the number of interaction steps matter.
This assumption is common, however, it is also somewhat unrealistic, especially in the contexts arising in robotics and embodied cognitive sciences\footnote{This is of course acknowledged by the community, and indeed, optimizing the computational complexity of useful learning algorithms is an important topic of research.}.
The first relaxation one can consider we call the \emph{weakly synchronous} settings.
In weakly synchronous settings, the interaction between the agent and the environment is still turn-based, but it takes place relative to an external clock. 
In static environments this does not change the situation. But in essentially any real setting, the environment does change and evolve, and it does so relative to the external clock (in \emph{real time}).
In this setting there are at least two types of benefits which stem from higher deliberation speeds, which is what quantization of model-based learning agents promises to provide. 
The benefit can be either trivial, or non-trivial.

By trivial, we refer to learning enhancements relative to a tester which samples from the interaction relative to the external clock, rather than the interaction steps. Here, if the model learns at all, higher speeds imply higher learning efficiency, relative to the externally-clocked tester.

Non-trivial cases pertain to, e.g. environments that change relative to the external clock. Here, faster deliberation times correspond to more slowly evolving environments, meaning the agent has more time to learn. To clarify, in the limit of ever more rapidly changing environments, the agent sees only noise, and there is nothing to learn at all. If the agent is extremely fast relative to the change, then the environment is effectively static, giving the agent ample time to learn.
Put in simple terms, we have shown that for model-based learning agents, in the settings where real time matters, quantum mechanics can essentially always help, merely by speeding-up the agent's deliberation times.
A more detailed investigation of quantum improvements of model-based learning will a topic of further research.
Regarding quantum advancements in weakly synchronous, classically communicating settings, it is worthwhile noting that works by other authors on quantum improvements in supervised learning, unsupervised learning, and other machine learning-related algorithmic tasks \cite{2013_Lloyd, 2013_Aimeur, 2012_Neven} may also significantly aid in learning from experience. Some of the initial approaches to reinforcement learning itself start from supervised and unsupervised learning techniques which are then tweaked to fit in the reinforcement learning paradigm \cite{SuttonBarto98}. It is likely that similar can be achieved for quantum algorithms as well, and it would be particularly interesting to see whether quantum access to environments would merge well with the quantum algorithms. In particular, quantum access to environments allows for the acquisition of superpositions of data, which is similar to the type of access achieved by employing Quantum Random Access Memory~\cite{2008_Giovannetti} utilized, for instance, in \cite{2013_Lloyd}. We leave such considerations for future investigation.

\section{Discussion}
\label{discussion}
In this work we have presented a framework for learning scenarios, where the agent-environment interaction can be fully quantum, which also allows the agent and the environment to become entangled.
The framework models the interaction as a sequence of CPTP maps, which the agent and the environment apply sequentially to their own register, and on the register(s) accessible to both parties. It is a fair question to ask to which extent this model is general.

 From the embodied agency perspective, an alternative formalism could posit local fields, specified by Hamiltonians, which the agent and the environment can switch on at will, and which act on the interfaces, and internal degrees of freedom of the agent and the environment. However, in the strictly synchronous model, the overall action is time-integrated, and amounts to one specified CPTP map. Thus the two approaches are equivalent. Extensions of our framework would have to commence already on the classical level, in which strict synchronicity,  discrete time or finiteness of action/percept spaces, would be dropped. 
 We have, implicitly, used some of such more general perspectives in the definition of the weakly synchronous model, and also through the use of scavenging and hijacking, which occur `between' legitimate time-steps. Any further generalization, with respect to timing, would have to consider dropping round-based processes, and should provide solutions to collisions, when the agent and environment produce continuous actions acting simultaneously on some of the registers. In this case, the Hamiltonian approach may prove to be more convenient. 
Regarding strictly synchronous models, which are standard in reinforcement learning, we have introduced the notion of a classical tester, as well as the sporadic tester, to enable a relatively straightforward comparison between classical agents and their quantum counterparts. However, the formalism allows for a large variety of testers, and it is possible that other types (e.g. testers which copy only the reward status, but not the actions or percepts) may lead to different types of improvements. 
It is not very likely that a notion of the tester can be completely circumvented, while still maintaining a reasonable notion of learning.
To argue this point further, note that in this work we posit the notion of the history of interaction as the central object, based on which what it means to learn is defined.
A meaningful (or useful) notion of history, in the quantum setting, should correspond to some observables (or sequences thereof) of the interface(s) of the agent and the environment, which is then mappable to a choice of a (perhaps somehow generalized) tester.
Alternatively, definitions of learning, which do not rely on a history, must then rely on the internal states of the agent which can be problematic\footnote{One can, for instance, imagine an agent that is (claims to be) learning, but refuses to demonstrate this. However, such an agent is operatively indistinguishable from an agent who is lying, or simply hallucinating.} and diverges from the standard notions of learning.

As the starting point of our approach, we have contrasted agent-environment interactions to oracular computational models, and have focused on techniques which, effectively, allow the agent to perform a Grover-type search on the environment. 
One of the advantages of this approach was that the statements about the performance could be given without an explicit reference to any specific learning model -- the examples we give start from any classical learning agent (essentially given as a black box) which are then generically enhanced by a quantum improvement in the exploration phase.
This is, arguably, the most straightforward, (and perhaps the most general) approach one can follow. 

However, we believe that even more striking results, which show a greater degree of separation between classical and quantum agents, are possible.
These could be, for instance, achieved by utilizing other types of oracles. In such a setting, we conjecture, the generality of the results is likely to become more modest, but may still include efficient solutions for interesting task environments. 

%A different approach towards , instead of `upgrading' a classical learning agent, take existing quantum algorithms for supervised and unsupervised learning, and modify them to realize a reinforcement learning agent. As such approaches were successful in classical AI, we expect that a similar paradigm may also be effective in a quantum setting.

 \noindent\textbf{Acknowledgments:\\}
VD and HJB acknowledge the support by the Austrian Science Fund (FWF) through 
the SFB FoQuS F 4012, and the Templeton World Charity Foundation grant TWCF0078/AB46.  VD thanks Christopher Portmann and Petros Wallden for useful discussions and comments on the manuscript.
\section{Appendix}
\label{appendix}

 \subsection{Proofs of Lemmas}
 In this section we give the proofs of the Lemmas from section \ref{QAEI}.\\
 
\noindent Proof of \textbf{Lemma \ref{LE-equiv}}.
\\
$(\Longrightarrow)$ If $A$ and $E$ have a classical interaction, by definition, at each stage of interaction, the state of the three registers $R_{A} R_C R_E$ is of the form 
\EQ{
\rho_{R_A R_C R_E} = \sum_{i} p_i\, \eta^i_{R_A} \otimes \rho^i_{R_C} \otimes \sigma^i_{R_E}.
} 
Moreover,  we have that ${\displaystyle \rho^i = \sum_{j \in \mathcal{S} \cup \mathcal{A}} q^i_j \dm{j}},$ for $q^i_j \in \mathbbmss{R}^+\cup \{ 0\}$ . Applying the classical tester yields the following state of $R_A R_C R_E R_T$:

\EQ{
\rho_{R_A R_C R_ER_T} =  \sum_{j \in \mathcal{S} \cup \mathcal{A}} \sum_{i}   p_i q^i_j\, \eta^i_{R_A} \otimes \dm{j}_{R_C} \otimes \sigma^i_{R_E} \otimes \dm{j}_{R_T},
} 
where we have, for clarity, commuted the sums over $i$ and over $j$.
It is now obvious that tracing out $R_T$ just recovers $\rho_{R_A R_C R_E},$ so this implication holds. \\
\vspace{0.5cm}

\noindent $(\Longleftarrow)$  We prove this direction by induction over interaction steps. Suppose the claim holds up to step $t-1,$ so at that time-step, the state of the three registers is 
 \EQ{
\rho^{t-1}_{R_A R_C R_E} = \sum_{i} p_i\, \eta^i_{R_A} \otimes \rho^i_{R_C} \otimes \sigma^i_{R_E}.
} 
Next, it is either the environmental or the agent's move. We will assume it is the agent's move, and the claim for the case of the environment's can be shown analogously.
The agent's map only sees registers $R_AR_C$ so we can write the state of the subsequent step as 
 \EQ{
\rho^{t}_{R_A R_C R_E} = \sum_{i} p_i\, \eta^i_{R_AR_C} \otimes \sigma^i_{R_E}.
} 
Now, each $ \eta^i_{R_AR_C}$ can be written as a convex combination of pure states:
\EQ{
\eta^i_{R_AR_C} = \sum_{j}{q^i}'_j {\dm{\psi_{i,j}}}_{R_AR_C},
} and each pure component $\ket{\psi_{i,j}}$ can be decomposed w.r.t. a separable basis:
\EQ{
\ket{\psi_{i,j}} = \sum_{k,l} \alpha_{k,l}^{i,j} \ket{\phi_{k}} \otimes \ket{x_{l}},
}
where $\ket{\phi_{k}}$ are classical states and $\ket{x_{l}}$ is a percept or an action state.
Putting it all together we have:
\EQ{
\eta^i_{R_AR_C} = \sum_{j}{q'}^i_j    \sum_{k,l,k',l'}      \alpha_{k,l}^{i,j}{\alpha_{k',l'}^{i,j}}^{\ast}\ket{\phi_k}\bra{\phi_{k'}}_{R_A} \otimes \ket{x_{l}}\bra{x_{l'}}_{R_C}
}
Copying the $R_C$ register (w.r.t. the classical basis), and then tracing out the copy system, reduces to eliminating all cross terms $ \ket{x_{l}}\bra{x_{l'}},$ where $l \not=l'.$ In other words, the following must hold, in order for the state to be invariant under classical testing:
 \EQ{
\sum_{i} p_i\,  \sum_{j}{q'}^i_j    \sum_{k,l,k',l'}      \alpha_{k,l}^{i,j}{\alpha_{k',l'}^{i,j}}^{\ast}\ket{\phi_k}\bra{\phi_{k'}}_{R_A} \otimes \ket{x_{l}}\bra{x_{l'}}_{R_C} =\\  \sum_{i} p_i\, \sum_{j}{q'}^i_j    \sum_{k,l,k',l'}      \alpha_{k,l}^{i,j}{\alpha_{k',l'}^{i,j}}^{\ast} \delta_{l,l'}\ket{\phi_k}\bra{\phi_{k'}}_{R_A} \otimes \ket{x_{l}}\bra{x_{l'}}_{R_C},
 }
where $\delta_{l,l'}$ is the Kronecker-delta.
So,
\EQ{
\rho^{t}_{R_A R_C R_E} =  \sum_{i}  \sum_{l} p_i\, \sum_{j}  \sum_{k,k'}  {q'}^i_j     \alpha_{k,l}^{i,j}{\alpha_{k',l}^{i,j}}^{\ast} \ket{\phi_k}\bra{\phi_{k'}}_{R_A} \otimes \ket{x_{l}}\bra{x_{l}}_{R_C} \otimes \sigma^i_{R_E},
}
 and by defining $\eta'_{l,i}=\sum_{j}  \sum_{k,k'}  {q'}^i_j     \alpha_{k,l}^{i,j}{\alpha_{k',l}^{i,j}}^{\ast}   \ket{\phi_k}\bra{\phi_{k'}} $ we get 
 \EQ{
 \rho^{t}_{R_A R_C R_E} =  \sum_{i} p_i\,   \sum_{l} {\eta'_{l,i}}_{R_A} \otimes \ket{x_{l}}\bra{x_{l}}_{R_C} \otimes \sigma^i_{R_E} = \sum_{i,l} p_i \,  {\eta_{l,i}}_{R_A} \otimes \dm{x_l}_{R_C} \otimes \sigma^i_{R_E}.
 }
% \blue{There must be a simpler way!!!}
The expression above is of the desired form, as soon as the sum is represented with one index.
Analogously, we obtain the claim for the environment's first move. This shows the step of the inductive proof, as the invariance under classical testing guarantees we always go from desired form states to desired form states.
To finish the inductive proof, we must establish the base of the induction. However, as we have clarified before, we assume that the initial state of the registers of the agent and environment is in product form, so this is trivial. \qed
\\ 
\vspace{1cm}

\noindent Proof of \textbf{Lemma \ref{LE-clas-int}}.
\\
This lemma essentially follows from the classical simulability of quantum mechanics. In particular, we can consider the classical agent $A^C$ and the classical environment $E^C$, which, internally, instead of storing quantum states, store the classical descriptions of the same quantum states: if the joint system, at time step $t-1$ of the registers $R_AR_CR_E$ is:
\EQ{
\rho_{R_AR_CR_E} = \sum_{i} p_i\, \eta^i_{R_A} \otimes \rho^i_{R_C} \otimes \sigma^i_{R_E}
 \label{quant}}
the corresponding state generated by $A^C$ and $E^C$ would be

\EQ{
\rho_{R_{A^C}R_CR_{E^C}} = \sum_{i} p_i\, [\eta^i]_{R_A} \otimes \rho^i_{R_C} \otimes [\sigma^i]_{R_E},
 \label{class}}
where with $[\rho]$ we denote the numerical matrix of the density operator $\rho$.
To clarify, the classical description $[\rho]$ of the quantum state $\rho$ is also a quantum state. However it is also always a classical state as $[\rho]$ and $[\rho']$ are orthogonal whenever $\rho \not= \rho'$. This may imply an exponential blow up in the number of registers needed, (and in the computation times), but this is irrelevant in the synchronous model of agent-environment interaction.

The transition to state $t$ is achieved by applying a map of the agent, or environment. Suppose it is the agent's move, as the argument will be analogous for the environmental move case.
At this point, the agent will apply a quantum map $\mathcal{M}$ to its system and the register $R_C,$ which maps 
\EQ{\eta^i_{R_A} \otimes \rho^i_{R_C} \stackrel{\mathcal{M}}{\longrightarrow} \sum_j q_j\   {\eta'}^j_{R_A} \otimes {\rho'}^j_{R_C},}
where the particular structure is ensured by the assumption the interaction is classical.
The classical agent can then be defined to apply a corresponding map mapping 

\EQ{[\eta^i]_{R_A} \otimes \rho^i_{R_C} \stackrel{[\mathcal{M}]}{\longrightarrow} \sum_j q_j\   [{\eta'}^j]_{R_A} \otimes {\rho'}^j_{R_C},
}
which is possible because the state $\rho^i$ is already a classical state, as the interaction is classical.
This establishes an inductive step. The basis of the induction also holds, provided that the initial state of the registers $R_AR_CR_E$ is a classical state, which, as we have clarified, we assume to be the case.
By inspecting equations $(\ref{quant})$ and  $(\ref{class})$ specifying the structure of the states of the registers realized by $A$ and $E$ and the classical counterparts $A^C$ and $E^C$ it is clear that the quantum histories generated by the two will be the same for all testers. \qed

\bibliographystyle{unsrt}

\end{document}